\documentclass[%
 reprint,
 amsmath,amssymb,
 aps,
 pra,
 floatfix,
]{revtex4-2}

\usepackage{graphicx}
\usepackage{dcolumn}
\usepackage{bm}
\usepackage[dvipsnames]{xcolor}

\usepackage{hyperref}
\hypersetup{
    colorlinks,
    linkcolor={red!50!black},
    citecolor={blue!50!black},
    urlcolor={blue!80!black}
}

\begin{document}

\preprint{APS/123-QED}

\title{Shortcuts to Adiabaticity in Anisotropic Bose-Einstein Condensates}

\author{Chinmayee Mishra}
 \email{chinmayee.mishra@oist.jp}
 \author{Thomas Busch}
 \email{thomas.busch@oist.jp}
 \author{Thom\'as Fogarty}
 \email{thomas.fogarty@oist.jp}
\affiliation{Okinawa Institute of Science and Technology, Okinawa, Japan}

\date{\today}

\begin{abstract}
    We propose shortcut to adiabaticity protocols for Bose–Einstein condensates trapped in generalized anisotropic harmonic traps in three dimensions. These protocols enable high-fidelity tuning of trap geometries on time scales much faster than those required for adiabatic processes and are robust across a wide range of interaction strengths, from weakly interacting regimes to the Thomas–Fermi limit. Using the same approach, we also design STA paths to rapidly drive interaction strengths in both isotropic and anisotropic traps. Comparisons with standard linear ramps of system parameters demonstrate significant improvements in performance. Finally, we apply these STA techniques to a unitary engine cycle with a BEC as the working medium. The STA methods significantly enhance the engine’s power output without reducing efficiency and remain highly effective even after multiple consecutive cycles.
\end{abstract}

\maketitle

\section{\label{sec:intro} Introduction}

Among the many techniques available for controlling and manipulating atomic Bose–Einstein condensates (BECs), two of the most widely used are the tuning of external harmonic trapping potentials~\cite{Dalfovo1999} and the variation of interaction strengths via magnetic Feshbach resonances~\cite{Chin2010}. However, unless these time-dependent changes in external control fields are performed adiabatically, they typically induce undesirable excitations in the condensate~\cite{Stringari1996, Perez1996, Andrews1997, Kagan1997, Polkovnikov2011, Cazalilla2010}. While slow adiabatic transformations can ensure that the final state is the stationary state of the system, their practical utility is often limited in quantum systems due to challenges such as atomic losses, decoherence, and other instabilities~\cite{Abeelen1999, Ketterle2002, Regal2004, Carvalho2004}.

To address this issue, modern non-adiabatic protocols have emerged, allowing the system to reach a desired target state within shorter time scales without generating unwanted excitations~\cite{Odelin2019}. These approaches, collectively termed shortcuts to adiabaticity (STAs), encompass a range of techniques that effectively mimic adiabatic transformations on much shorter time scales. This is achieved by engineering alternative paths for varying the control fields, which often deviate significantly from the conventional linear or monotonic variations used in the adiabatic limit. Prominent methods include inverse engineering~\cite{Torrontegui2014, Vitanov2015, Zhang2017, Impens2017}, counter-diabatic driving~\cite{Berry2009, Campo2013}, invariant-based engineering~\cite{Chen2010}, variational methods~\cite{Li2016, Li2018}, and the fast-forward approach~\cite{Masuda2008}. The effectiveness of these methods has been demonstrated in several notable experimental studies across a variety of quantum systems~\cite{Schaff2010, Schaff2011, Bowler2012, Rohringer2015, Deng2018}.

While most early studies were focused on linear or single-particle systems, recent efforts have extended the concept of STAs to more complex few- and many-particle systems \cite{Campo2012, Beau_2016, Sels2017, Skycow2020, Fogarty_2021, COLD_2023, Hasan2024, Morawetz_2024}. One example is interacting atomic BECs, which in the mean-field limit can be described by a single non-linear field equation. However, the presence of the non-linearity poses significant challenges to finding efficient STA protocols, as it renders many established techniques ineffective. To date, most studies of STAs for BECs have focused on symmetrically trapped condensates, where their self-similar dynamics can often be captured by a single equation of motion \cite{Deffner2014, Muga2009}. Engineering STAs for BECs for anisotropic structural changes in the confining geometry remains challenging, as coupling between different degrees of freedom must be controlled. To address this, we here adopt an effective scaling approach to describe the evolution of the density distribution, which is an approach that has proven effective in modeling expansions in Bose gases \cite{Odelin2002}, Fermi gases \cite{Menotti2002}, and Bose-Fermi mixtures \cite{Hu2003}. It utilizes a self-similar ansatz to capture the system’s evolution in the hydrodynamic regime and, unlike the pure scaling approach, which requires the scaling to be an exact solution of the hydrodynamic equations, the effective scaling approach remains valid as long as an approximate self-similarity can be sustained \cite{Modugno2018, Viedma2020}.


In the following, we employ the effective scaling approach to demonstrate the possibility to inverse engineer STAs for BECs in anisotropic harmonic traps, covering the full interaction range from the weakly interacting regime to the Thomas–Fermi limit. We design STAs that allow for individual control of trap frequencies in three dimensions, demonstrating their effectiveness for arbitrary trap ratios and geometries. For instance, we show that a BEC can be deformed from an isotropic to a cigar-shaped density profile with high fidelity in finite time across a broad range of interaction strengths. Additionally, we extend the effective scaling approach to design STAs for controlling the interactions in a BEC in a fixed anisotropic trapping potential, however we find that the STAs underperform when  comparing to similar interaction STAs designed using a variational approach. 


A key application of STAs lies in enhancing the performance of quantum engines, where rapid implementation of individual engine strokes, combined with the suppression of non-adiabatic excitations, results in increased power output \cite{Campo2014-cf, Campo2016}. Recent theoretical \cite{Myers2022, Estrada2024} and experimental \cite{Simmons2023} studies have investigated quantum engines using BECs as the working medium, and have demonstrated that operating in the condensate regime can enhance performance. Inspired by the experiment reported in Ref.~\cite{Simmons2023}, we apply our STA protocols to a comparable isentropic engine cycle. Our results show that these protocols enable enhanced power output without compromising engine efficiency, and this performance can be sustained over multiple consecutive cycles.

The manuscript is organized as follows. In Sec.~\ref{sec:scaling}, we introduce the BEC system and outline the effective scaling self-similar approach for a generalized anisotropic three-dimensional system. We derive the hydrodynamic equations, which lead to effective Ermakov-like equations, and in Sec.~\ref{sec:equations}, these equations are employed to develop STAs for the expansion and compression of the BEC. We compare our generalized approach to STAs using the geometric mean, demonstrating significant improvements in performance using our protocols. We then develop STA protocols to modify interaction strengths and show that the previously adopted self-similar scaling approach performs poorly for anisotropically trapped BECs. To address this, we develop an STA using a variational approach within the Thomas-Fermi regime. After formulating STAs for both trap and interaction ramps, we apply these anisotropic shortcuts in Section~\ref{sec:engine} to model an isentropic engine cycle~\cite{Simmons2023}, achieving excellent fidelity and enhanced power output. Finally, Section~\ref{sec:conclusion} summarizes our findings and discusses potential extensions of our model.

\begin{figure}[t]
    \centering
    \includegraphics[width=\linewidth]{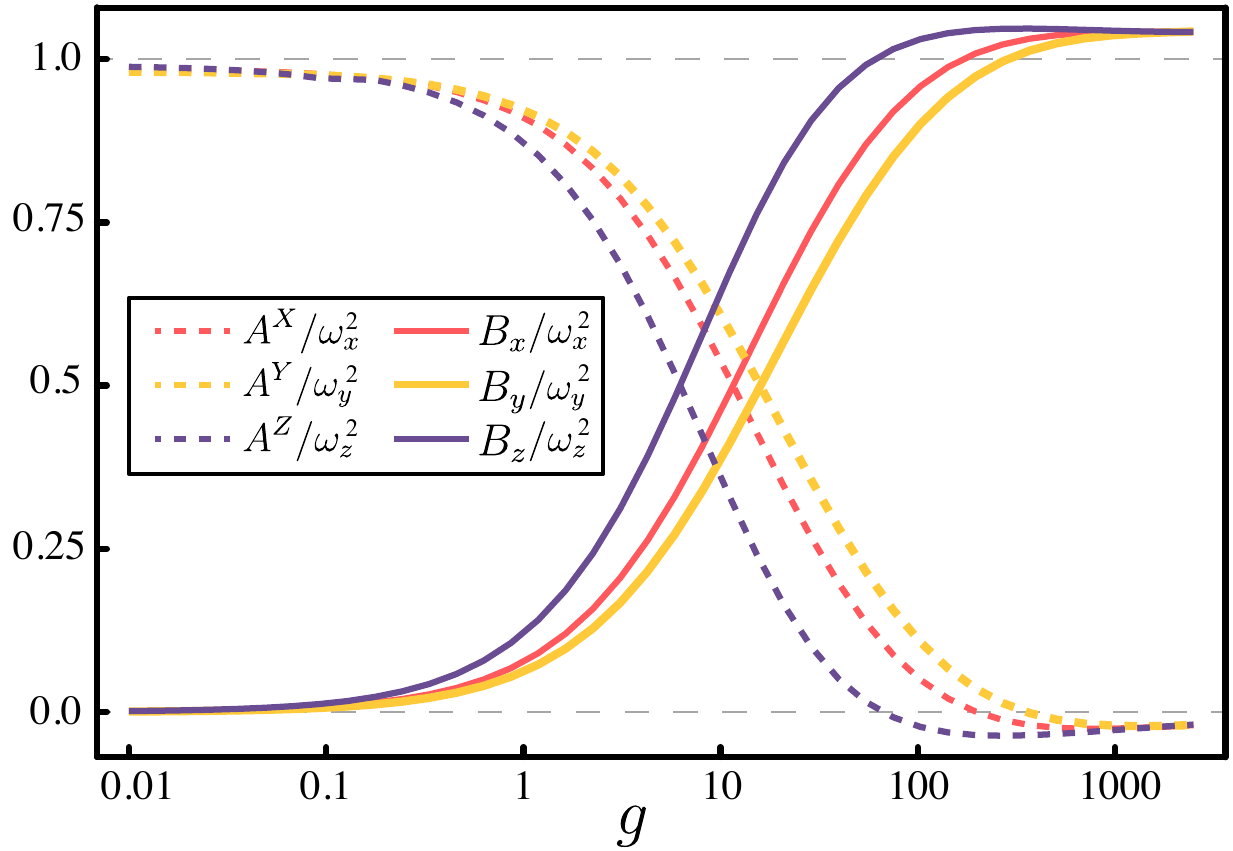}
    \caption{The coefficients $A^\sigma/\omega_\sigma^2 = (A^\sigma_x + A^\sigma_y + A^\sigma_z)/\omega_\sigma^2$ and $B_\sigma/\omega_\sigma^2$ 
    as functions of the interaction strength $g$. The condition $(A^\sigma + B_\sigma)/\omega_\sigma^2 = 1$ is satisfied for $(\omega_x, \omega_y, \omega_z) = (0.8, 1, 0.5)\omega_0$. 
    }
    \label{fig:abcoeff}
\end{figure}

\section{\label{sec:scaling} Effective Scaling Approach}
A three-dimensional atomic BEC confined in a harmonic trap can be accurately described in the weakly correlated regime using a mean-field approach. In this framework, the condensate is represented by a macroscopic wavefunction, $\Psi(\mathbf{r},t)$, governed by the non-linear Gross–Pitaevskii equation (GPE) \cite{PethickSmith2008}
\begin{align}
    i \dfrac{\partial}{\partial t}\Psi(\mathbf{r},t)= \left[-\dfrac{\nabla^2}{2}+V(\mathbf{r},t)+g(t)|\Psi(\mathbf{r}, t)|^2\right]\Psi(\mathbf{r},t)\;.\label{eq:gpe}
\end{align}
The harmonic potential is expressed in its general form as  $V(\mathbf{r},t) = \left(\omega_x(t)^2 x^2 + \omega_y(t)^2 y^2 + \omega_z(t)^2 z^2\right)/2$ , where the trap frequencies $\omega_{\sigma\in(x,y,z)}$ may vary over time and are not necessarily equal. The short-range interaction strength is denoted by $g$, which is also allowed to vary with time.

For ease of notation, we scale all lengths by $\sqrt{\hbar/m\omega_0}$, time by $\omega_0^{-1}$, and energy by $\hbar\omega_0$, rendering Eq.~\eqref{eq:gpe} dimensionless. Here, $\omega_0$ is an arbitrary reference frequency chosen for convenience. Additionally, the normalization condition $\int d^3\mathbf{r}|\Psi(\mathbf{r}, t)|^2 = 1$ is maintained at all times.

The corresponding hydrodynamic equations for the density, $n(\mathbf{r}, t) = |\Psi(\mathbf{r}, t)|^2$, and the velocity field, $\mathbf{v} = (\Psi^*\nabla\Psi - \Psi\nabla\Psi^*)/(2i|\Psi|^2)$, can be derived by formulating the Lagrangian of the system and applying the Euler–Lagrange equations of motion \cite{Viedma2020, Huang2021, Modugno2018}
\begin{align}
    &\dfrac{dn(\mathbf{r}, t)}{dt} + \nabla\cdot[n(\mathbf{r}, t)\mathbf{v}] = 0\;,\nonumber\\
    &\dfrac{d\mathbf{v}}{dt} + \nabla\cdot\left[P(\mathbf{r}, t)+V(\mathbf{r}, t)+\dfrac{1}{2}\mathbf{v}^2+g(t)n(\mathbf{r}, t)\right] = 0\;,\label{eq:hydrodyanmic}
\end{align}
where $P(\mathbf{r}, t) = -\nabla^2\sqrt{n(\mathbf{r}, t)}/(2\sqrt{n(\mathbf{r}, t)})$ is the so-called quantum pressure.  These coupled equations govern the dynamics of the condensate. To derive the coupled  equations of motion for the system in the different spatial directions, we now employ a scaling solution of the form $n = n_0(x/x_0, y/y_0, z/z_0)/(x_0y_0z_0)$, where $x_0(t)$, $y_0(t)$ and $z_0(t)$ are dimensionless scaling parameters that characterize the time-dependent width of the condensate, and $v_{\sigma\in(x, y, z)} = \dot{\sigma}_0 \sigma / \sigma_0$.

Following the algebraic procedure detailed in Appendix \ref{app:3dermakov}, this leads to a set of three coupled equations for the scaling parameters of the form
\begin{subequations}
\begin{align}
    \ddot{x}_0+\omega_x^2(t)x_0 = \dfrac{A_x^x}{x_0^3} + \dfrac{A_y^x}{x_0y_0^2} + \dfrac{A_z^x}{x_0z_0^2} + \dfrac{B_x}{x_0^2y_0z_0},\label{eq:trap3der1}\\
    \ddot{y}_0+\omega_y^2(t)y_0 = \dfrac{A_x^y}{y_0x_0^2} + \dfrac{A_y^y}{y_0^3} + \dfrac{A_z^y}{y_0z_0^2} + \dfrac{B_y}{x_0y_0^2z_0},\label{eq:trap3der2}\\
    \ddot{z}_0+\omega_z^2(t)z_0 = \dfrac{A_x^z}{z_0x_0^2} + \dfrac{A_y^z}{z_0y_0^2} + \dfrac{A_z^z}{z_0^3} + \dfrac{B_z}{x_0y_0z_0^2},\label{eq:trap3der3}
\end{align}
\end{subequations}
where for $\sigma, \sigma' \in(x, y, z)$ 
\begin{align}
    A_{\sigma'}^\sigma = \dfrac{D^{\sigma=0}_{\sigma'}-E_{k}^0(\sigma)}{E_{v}^0(\sigma)};\qquad B_\sigma = \dfrac{g(n_0^{\sigma=0}-E_{int}^0)}{E_{v}^0(\sigma)},\nonumber
\end{align}
and
\begin{align}
  P_\sigma &= -\nabla^2_\sigma\sqrt{n_0}/2\sqrt{n_0},\\ D_{\sigma'}^{\sigma=0}&=\int P_{\sigma'}(R, t)|_{\sigma=0} n_0(R, t)d^3R,\\
  n_0^{\sigma=0}&=\int n_0(R, t)|_{\sigma=0}n_0(R, t)d^3R.
\end{align}  
Here, $R^2 = X^2 + Y^2 + Z^2$, with the scaled lengths $(X, Y, Z) = (x, y, z) / (x_0, y_0, z_0)$. The kinetic energy is given by $E_\text{kin}^0(\sigma) = \int P_\sigma n_0 d^3R$, the potential energy by $E_\text{pot}^0(\sigma) = \int \sigma^2 n_0 d^3R / 2$, and the interaction energy by $E_\text{int}^0 = \int n_0^2 d^3R$. The coefficients in Eqs.~\eqref{eq:trap3der1}–\eqref{eq:trap3der3} satisfy $\sum_{\sigma’} (A_{\sigma’}^\sigma + B_\sigma) / \omega_\sigma^2(0) = 1$, which can be demonstrated by setting $\Psi = \sqrt{n_0}$ and the left-hand side of Eq.~\eqref{eq:gpe} to $\mu_\sigma \sqrt{n_0}$ at $t=0$. Additionally, we note that $x_0 = y_0 = z_0 = 1$ at $t=0$. Left-multiplying by $\sqrt{n_0}|_\sigma$ then yields an expression for $\mu\sigma$ and substituting this into the original equation confirms that the coefficients sum to one (see also Appendix~\ref{app:3dermakov}). In Fig.~\ref{fig:abcoeff}, we plot the coefficients over a range of interaction strengths, from weak to strong interactions.

\begin{figure*}[t]
    \centering
    \includegraphics[width=\linewidth]{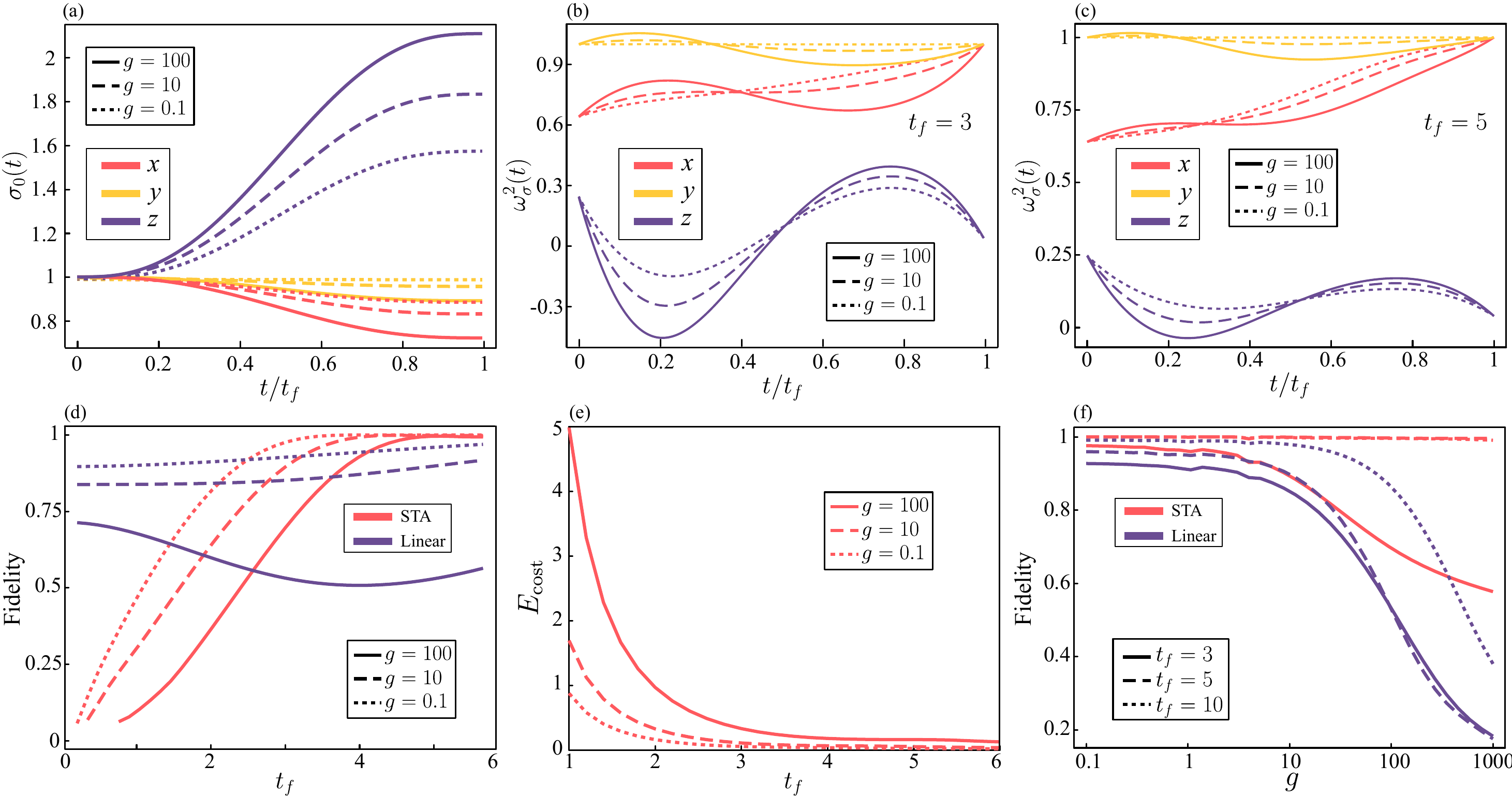}
    \caption{
    Anisotropic trap ramp using shortcuts to adiabaticity (STA) from a 3D initial trap $(\omega_x, \omega_y, \omega_z) = (0.8, 1.0, 0.5)$ to a cylindrical trap $(1.0, 1.0, 0.2)$. (a) The scaling parameters along the three spatial directions are shown for three different $g$ values, directly obtained using Eq.~\eqref{eq:polynomial}. (b, c) The shortcut paths for $\omega_x$, $\omega_y$, and $\omega_z$ are illustrated for $t_f = 3$ and $t_f = 5$, corresponding to the specified $g$ values. (d) Fidelity comparison between STA and a linear ramps. STA ramps quickly achieve high-fidelity states, outperforming linear ramps. (e) The energy cost associated with STA ramps, corresponding to the fidelity results in the previous subfigure. (f) Fidelity comparison between STA and linear ramps as a function of $g$, plotted on a logarithmic scale for different $t_f$ values. Notably, the red dashed and dotted lines overlap, indicating that the STA achieves perfect fidelity for times above $t_f = 5$.}
    \label{fig:trapramp}
\end{figure*}

In the limiting case of $g = 0$, the $B_\sigma$ terms vanish~\cite{Huang2021, Modugno2018}, and the cross terms $A_{\sigma’}^\sigma$ also disappear. This leaves the anisotropic 3D Ermakov equations for a non-interacting gas
\begin{align}
    \ddot{\sigma}_0 + \omega_\sigma^2(t)\sigma_0 = \frac{\omega_\sigma(0)^2}{\sigma_0^3},
\end{align}
where, as expected, the three equations are decoupled and thus behave independently in the three spatial directions. Conversely, in the Thomas–Fermi limit of strong interactions, all $A_{\sigma’}^\sigma$ approach zero, and $B_\sigma \approx \omega_\sigma^2(0)$. Under these conditions, the equations reduce to the three-dimensional Ermakov-like equations for a Thomas–Fermi condensate in an anisotropic harmonic trap given by, 
\begin{subequations}    
\begin{align}
    \ddot{x}_0+\omega_x^2(t)x_0 &= \dfrac{\omega_x(0)^2}{x_0^2y_0z_0}\,,\\
    \ddot{y}_0+\omega_y^2(t)y_0 &= \dfrac{\omega_y(0)^2}{y_0^2z_0x_0}\,,\\
    \ddot{z}_0+\omega_z^2(t)z_0 &= \dfrac{\omega_z(0)^2}{z_0^2x_0y_0}\,.
\end{align}
\end{subequations}
Clearly, in the isotropic scenario, the equations reduce to the well-known result $\ddot{\sigma}_0+\omega_\sigma^2\sigma_0^2=1/\sigma_0^4$ \cite{Muga2009}. These equations are strongly coupled but remain independent of the interaction strength. In Fig.~\ref{fig:abcoeff}, for strong interactions, when the coefficients take values $A_{\sigma’}^\sigma < 0$ and $B_\sigma > 1$ 
, a decrease in fidelity might be expected as the scaling approach begins to break down, as shown in \cite{Viedma2020, Huang2021}. However, this effect is negligibly small in our system. Furthermore, we show that in this regime the fidelity can be significantly improved by slightly increasing the duration of the shortcut, as we will discuss in the following section.

\section{\label{sec:equations} Shortcuts to Adiabaticity}

In this section, we utilize the Ermakov-like Eqs.~\eqref{eq:trap3der1}–\eqref{eq:trap3der3} to develop  frictionless trajectories for varying system parameters, such that a desired final state can be reached in significantly shorter times compared to adiabatic time scales. These shortcut paths specify the time-dependent tunings of the various Hamiltonian parameters, namely the trapping frequencies and the interaction strength. In the following we will first systematically derive the shortcut paths for varying the trap frequencies in the three spatial directions while maintaining a fixed interaction strength. After that, we will derive STAs for altering the interaction strength under a fixed confinement geometry.

\subsection{Anisotropic Trap Ramps}
The objective is to vary the trap frequencies from $\omega_{\sigma}(0)$ to their final values $\omega_\sigma(t_f)$ within the total time $t_f$, without inducing unwanted excitations in the condensate. Such processes have practical applications, for instance, in frictionless cooling when $\omega_\sigma(t_f)/\omega_\sigma(0) < 1$, which has been theoretically studied across various interaction regimes in isotropic traps \cite{Muga2009, Chen2010, Huang2020, Huang2021}. Our approach generalizes this concept to gases confined in anisotropic traps, accommodating any interaction regime and both compression and expansion of the trap frequencies.

To achieve this, it is desirable for the scaling parameters $\sigma_0$ to remain stationary at both the beginning and end of the trap ramp. Specifically, we impose the boundary conditions $\dot{\sigma}_0(0) = \ddot{\sigma}_0(0) = \dot{\sigma}_0(t_f) = \ddot{\sigma}_0(t_f) = 0$, and substituting these conditions into Eqs.~\eqref{eq:trap3der1}–\eqref{eq:trap3der3} for a fixed $g$, we obtain
\begin{align}
    A_x^x y_0^2 z_0^2 + A^x_y z_0^2 x_0^2 + A_z^x x_0^2 y_0^2 + B_x x_0 y_0 z_0 = \omega_x^2x_0^3y_0^2z_0^2\,,\nonumber\\
    A_x^y y_0^2 z_0^2 + A^y_y z_0^2 x_0^2 + A_z^y x_0^2 y_0^2 + B_y x_0 y_0 z_0 = \omega_x^2x_0^3y_0^2z_0^2\,,\nonumber\\
    A_x^z y_0^2 z_0^2 + A^z_y z_0^2 x_0^2 + A_z^z x_0^2 y_0^2 + B_z x_0 y_0 z_0 = \omega_x^2x_0^3y_0^2z_0^2\,.\label{eq:bv}
\end{align}
Simultaneously solving the coupled equations above for $\omega_\sigma$ at the known initial and final times provides the boundary conditions for $\sigma_0$ at $t = 0$ and $t = t_f$, respectively.

We adopt a polynomial ansatz for the trajectory of the scaling parameters, ensuring that the initial and final states are stationary. This ansatz takes the specific form
\begin{align}
    \sigma_0(t) = \sigma_0(0) - 6[\sigma_0(0)-\sigma_0(t_f)](t/t_f)^5\,\, \nonumber\\
    + 15[\sigma_0(0)-\sigma_0(t_f)](t/t_f)^4\,\, \nonumber\\
    - 10[\sigma_0(0)-\sigma_0(t_f)](t/t_f)^3\,,\label{eq:polynomial}
\end{align}
which depends only on $t/t_f$ and is therefore applicable for any value of $t_f$. In Fig.~\ref{fig:trapramp}, we show an example of a ramp transitioning from $(\omega_x, \omega_y, \omega_z) = (0.8, 1, 0.5)$ to $(1, 1, 0.2)$. Panel (a) depicts the scaling parameters, which exhibit smooth and monotonic behavior.

Substituting this ansatz into Eqs.\eqref{eq:trap3der1}–\eqref{eq:trap3der3} determines the shortcut paths for evolving $\omega_\sigma(t)$ over the duration $t_f$. These paths are plotted for two values of $t_f$ in Figs.~\ref{fig:trapramp}(b)–(c). Notably, due to the coupling between the three spatial directions imposed by interactions, the trap frequency in the $y$ direction must also be modulated during the STA, even though its initial and final values remain unchanged, $\omega_y(0) = \omega_y(t_f)$. Furthermore, stronger interaction strengths result in greater coupling between spatial dimensions, necessitating more significant modulation of the trap frequencies to achieve the STA. It is important to note that the trap frequencies can become negative for rapid driving times \cite{Huang2021}, corresponding to an inversion of the confining potential. This effect, observed for example in $\omega_z^2(t)$, can impose practical limitations on driving times in experiments \cite{Campo2014-cf, Abah2018, Zhang2023-mx}. However, we emphasize that trap inversion alone does not lead to instability in the BEC, provided the scaling approach remains an exact description of the system.

To quantify the effectiveness of the STA, we calculate the fidelity, $\mathcal{F} = |\langle\Psi(t_f)|\Psi_\text{tar}\rangle|^2$, which measures the overlap between the state at the end of the driving process, $\Psi(t_f)$, and the target state, $\Psi_\text{tar}$, corresponding to the ground state of the BEC with the final trapping frequency $\omega_f$. As shown in Fig.~\ref{fig:trapramp}(d), our STA can achieve perfect fidelities and significantly outperforms a linear ramp for intermediate timescales which are comparable to the inverse trap frequency. However, it is important to note that the STA is not effective for very short driving times due to the limitations of our ansatz in accurately describing the dynamics of a BEC with finite nonlinear coupling. The large variations in trap frequencies over short timescales drive the system far out of equilibrium, which is a well-known limitation of approximate STAs. Nonetheless, the STA consistently outperforms the linear ramp, particularly in the strongly interacting regime where the dynamics in the three spatial directions are strongly coupled. In this regime, longer driving times are required to achieve unit fidelity, as the rapid modulation of trapping frequencies necessary for quickly controlling strongly interacting BECs can induce significant irreversible excitations. This behavior is also related to the region in Fig.~\ref{fig:abcoeff}, where $A_{\sigma’}^{\sigma} < 1$ and $B_\sigma > 1$ \cite{Modugno2018, Viedma2020}. Despite this, the STA remains effective and continues to outperform linear ramps.


The energetics of the STA plays a crucial role in achieving high-fidelity control of the BEC. This can be quantified through the energetic cost of the STA ramp \cite{Li2018, Obinna2019, Obinna2019_2, Obinna2020},
defined as
\begin{align}
    E_{\text{cost}} = \frac{1}{t_f} \int_0^{t_f} dt , [\epsilon_{\text{STA}}(t) - \epsilon_{\text{ad}}(t)],
\end{align}
where $\epsilon_{\text{STA}}(t)$ represents the instantaneous energy at time $t \in [0, t_f]$ during the STA ramp, and $\epsilon_{\text{ad}}(t)$ corresponds to the stationary energy for an adiabatic ramp. The instantaneous energy, $\epsilon(t)$, is given by
\begin{align}
    \epsilon_{\text{STA}}(t) = \int d\textbf{r} \Psi^*\left[-\dfrac{\nabla^2}{2}+V(\textbf{r}, t)+\dfrac{g(t)}{2}|\Psi|^2\right]\Psi\,.
\end{align}
In Fig.~\ref{fig:trapramp}(e), we show that STAs generally incur a higher energy cost for stronger interactions compared to their respective adiabatic energies. This observation is consistent with the need for longer times when applying an STA to a strongly interacting BEC to achieve perfect fidelity. Predictably, at short timescales, the energy cost increases for all interaction strengths as more energy is required to drive the system at higher speeds. By choosing sufficiently longer durations for $t_f$, while still remaining much shorter than adiabatic timescales, this slight disadvantage can be mitigated, enabling perfect fidelity across a wide range of $g$ values, as shown in Fig.~\ref{fig:trapramp}(f). In contrast, linear ramps fail to reach even half the maximum fidelity in the strong interaction regime.

\begin{figure}
    \centering
    \includegraphics[width=\linewidth]{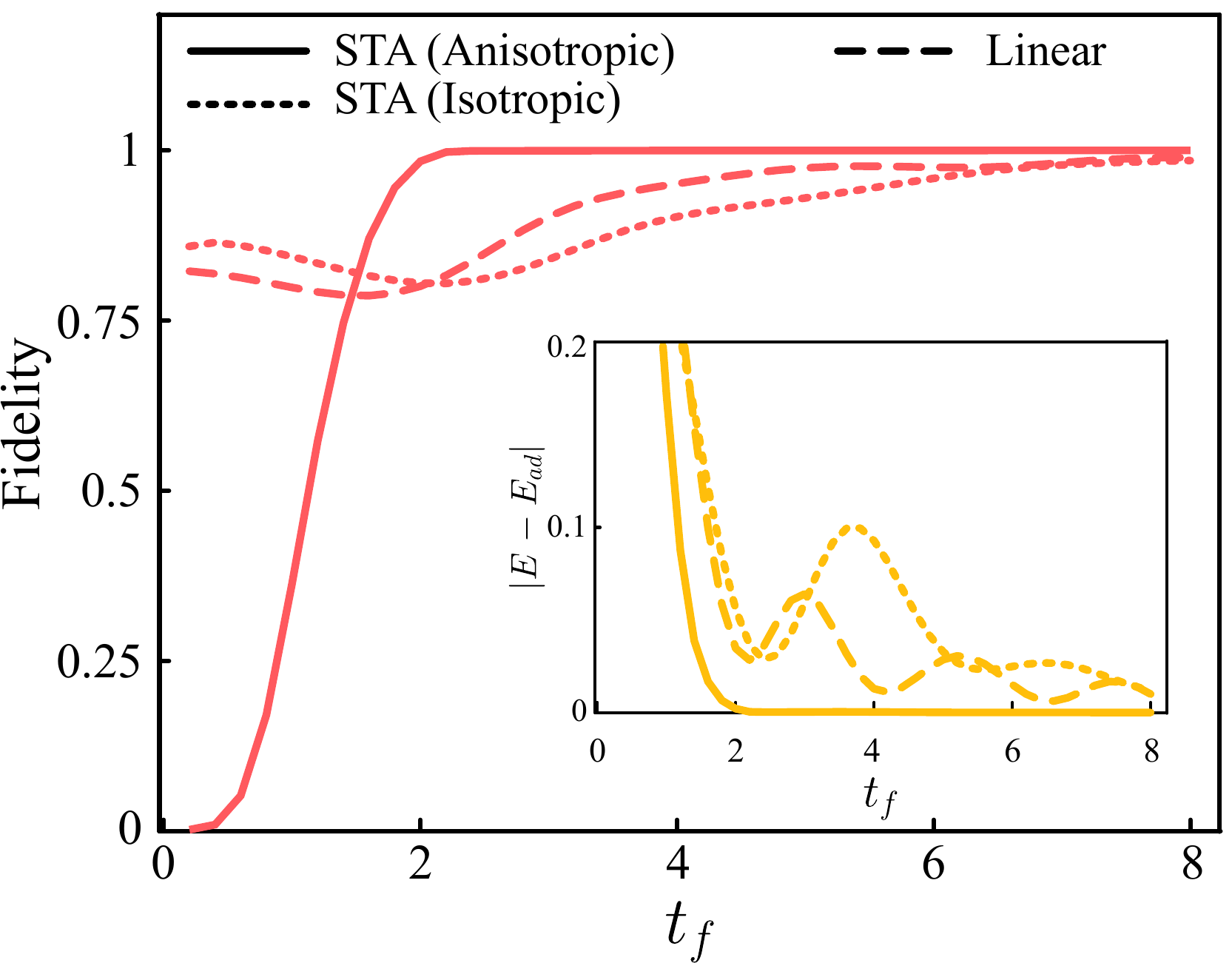}
    \caption{Fidelity comparison between two different STA ramps and a linear ramp for a slightly anisotropic trapped BEC. The initial and final traps are $(\omega_x, \omega_y, \omega_z) = (0.3, 0.5, 0.4)$ and $(2\omega_x, 2\omega_y, 2\omega_z)$, respectively, at a fixed interaction strength $g = 100$. Inset: Energy difference between the adiabatic energy and the energy of the dynamically evolved final state.
    }
    \label{fig:compare}
\end{figure}

As discussed above, the STA protocol relies on the independent control of the three trapping frequencies, which increases the complexity of the problem. For this reason we next check the performance of a simpler approach  that designs an STA based on the geometric mean of the trapping frequencies, $\bar{\omega}(t) = (\omega_x(t)\omega_y(t)\omega_z(t))^{1/3}$. In this method, the individual trap frequencies are given by
\begin{equation}
    \omega_{\sigma}(t) = \omega_{\sigma}(0) + \left( \frac{\bar{\omega}(t) - \bar{\omega}(0)}{\bar{\omega}(t_f) - \bar{\omega}(0)} \right)\left(\omega_{\sigma}(t_f) - \omega_{\sigma}(0)\right),
\end{equation}
where the modulation of each trapping frequency is governed by the same time-dependent function, $\bar{\omega}(t)$. 
In the Thomas–Fermi (TF) limit, the relevant Ermakov equation \cite{Muga2009} governing isotropic 3D traps is
\begin{align}
    \ddot{\sigma}_0+\bar{\omega}(t)^2\sigma_0 = \dfrac{\bar{\omega}(0)^2}{\sigma_0^4}\;, \label{eq:iso}
\end{align}
and to evaluate the performance of this approach, we compare fidelities achieved using the full anisotropic STA from Eqs.~\eqref{eq:trap3der1}–\eqref{eq:trap3der3}, the simplified geometric mean STA from Eq.~\eqref{eq:iso}, and linear ramps of the trap. 

In Fig.~\ref{fig:compare}, we present fidelity curves for the three trap ramps, where the trap frequencies are doubled without altering the anisotropy ratios, i.e., $\bar{\omega}(t_f) = 2\bar{\omega}(0)$. One can clearly see that even for traps with low anisotropy, the difference between anisotropic and geometric mean STAs is significant. This contrast becomes especially evident when examining the energy difference between the target state and the state at the end of the ramp, $|E - E_\text{ad}|$, as shown in the inset of Fig.~\ref{fig:compare}. While the anisotropic STA successfully reaches the target energy, the geometric mean STA introduces significant excitations, particularly for times shorter than $t_f \sim 3.8$. This highlights the importance of independent control over the trapping frequencies in all three spatial directions for efficiently driving anisotropic BECs.

\begin{figure}[t]
    \centering
    \includegraphics[width=\linewidth]{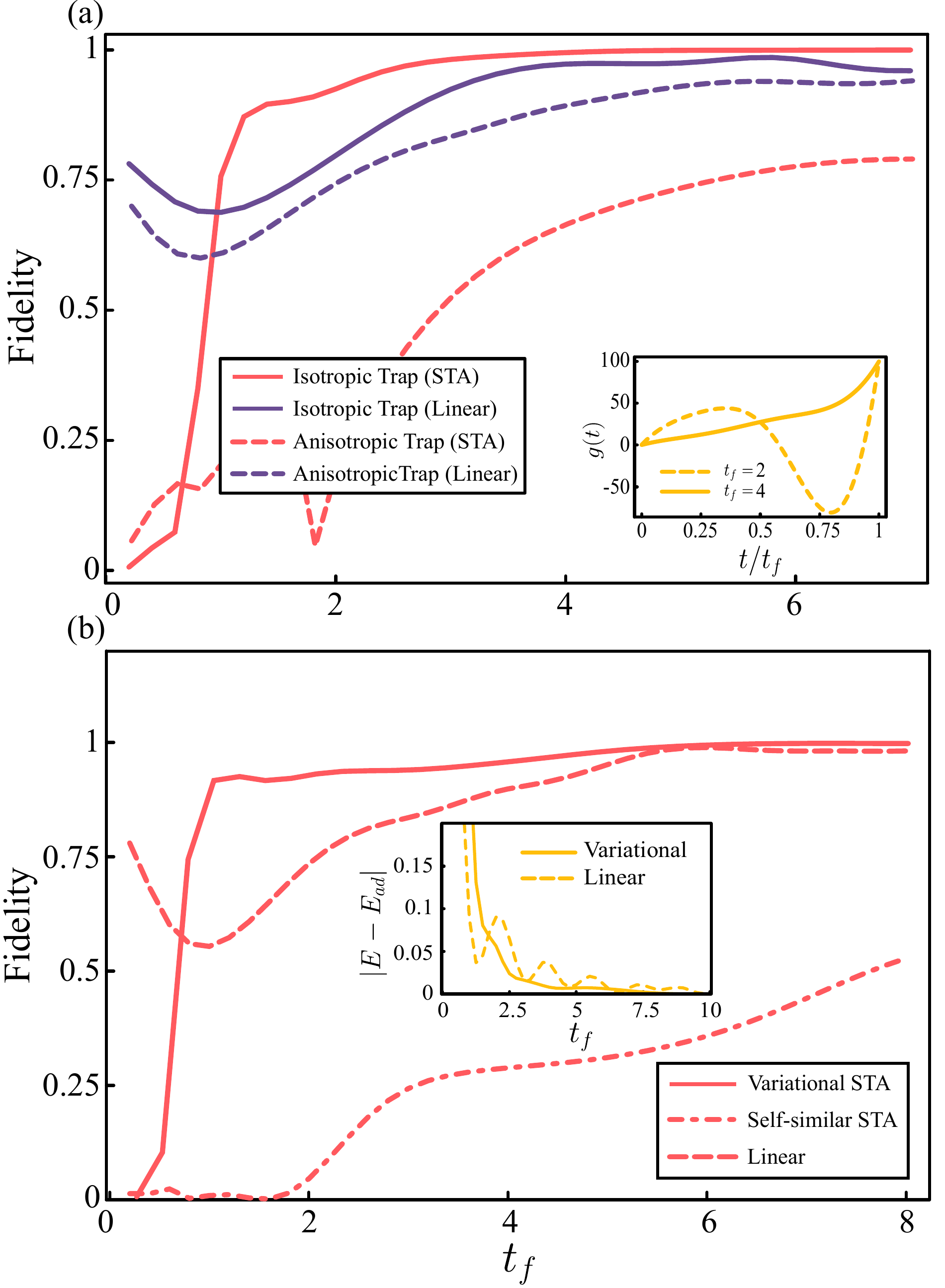}
    \caption{(a) Fidelity comparison between STA and linear ramps of the interaction, varying from $g = 0.1$ to $g = 100$, for an isotropic 3D trap $(\omega_x, \omega_y, \omega_z) = (0.6, 0.6, 0.6)$ and an anisotropic trap $(\omega_x, \omega_y, \omega_z) = (0.8, 1.0, 0.5)$. The inset shows the shortcut paths for the isotropic trap case for two different $t_f$ values. (b) Fidelity comparison between different ramps when the interaction strength is varied from $g = 100$ to $g = 500$ in the same anisotropic trap as above. The inset shows the energy difference between the final state and the target adiabatic state.
    }
    \label{fig:gramp}
\end{figure}

\subsection{Interaction Ramp}
The same Ermakov-like equations, Eqs.~\eqref{eq:trap3der1}-\eqref{eq:trap3der3}, can also be applied to determine shortcut paths for interaction ramps under fixed trapping frequencies. Following a similar procedure as for the trap ramps, we redefine $B_\sigma = g(t)B_\sigma’$, fixing the interaction strength at $t = 0$ and $t = t_f$ as $g(0)$ and $g(t_f)$, respectively. This leads to Eq.~\eqref{eq:polynomial}, which is then substituted into the Ermakov-like equations leading to the expression
\begin{align}
    g(t) = \dfrac{x_0y_0z_0}{B'}\sum_{\sigma\in(x, y, z)}\left(\ddot{\sigma}_0\sigma_0+\omega_\sigma^2\sigma_0^2-\dfrac{A_\sigma}{\sigma_0^2}\right)\label{eq:gt},
\end{align}
where $B'=\sum_{\sigma\in(x, y, z)}B'_\sigma$ and $A_{\sigma\in(x, y, z)} = \sum_{\sigma'} A^{\sigma'}_\sigma$.

In Fig.~\ref{fig:gramp}(a), we show the case where the interaction strength is increased from the weakly interacting regime to the TF limit for two different trap geometries: an isotropic and an anisotropic trap. One can see that the STA protocol successfully achieves high fidelities in the isotropic trap and outperforms linear ramps for $t_f > 1$. The inset shows STA ramps for the isotropic case with two different $t_f$ values. For faster ramps, the STA requires the interaction to briefly switch from repulsive to attractive, an effect similar to the trap inversion discussed earlier. On the other hand, for anisotropic traps, the STA is significantly less effective, as drastic variations in the interactions are poorly captured by the effective scaling approach \cite{Viedma2020}. In fact, in these situations the STA performs even worse than a linear ramp.

To improve this limited performance, we will use a variational method~\cite{Huang2020} and show that it gives good results, particularly when operating within the Thomas–Fermi limit. Additionally, switching from repulsive to attractive interactions during the intermediate part of the ramp can be largely avoided in this regime. For this we start with a variational ansatz of the form
\begin{align}
    \Psi=\sqrt{\frac{15}{8\pi a_xa_ya_z}}\sqrt{1-\frac{x^2}{a_x^2}-\frac{y^2}{a_y^2}-\frac{z^2}{a_z^2}} e^{i(x^2c_x+y^2c_y+z^2c_z)}\,
\end{align}
where the parameters $a_\sigma$ are the scaling lengths in the three spatial directions. However, the three different scaling lengths can be reduced to a single one by using the approximation $a_\sigma = b(t)/\omega_\sigma$ in the Thomas--Fermi regime. Substituting this into the Lagrangian density $\mathcal{L} = i(\Psi\dot{\Psi}^*-\Psi^*\dot{\Psi})/2-|\nabla\Psi|^2/2 -V(r,t)-g|\Psi|^4$ and following the steps in \cite{Perez1996} we obtain the expression
\begin{align}
    \ddot{b}b\sum_\sigma \dfrac{1}{\omega_\sigma^2}+3b^2 = \dfrac{35\log 2\sum_\sigma \omega_\sigma^2}{2b^2}+\dfrac{45g(t)\prod_\sigma\omega_\sigma}{4\pi b^3},\label{eq:gtvar}
\end{align}
from which the expression $g(t)$ can be extracted via inverse engineering as done previously for the trap ramps. We present the results for varying the interactions purely within the TF regime, transitioning from $g = 100$ to $g = 500$, in Fig.~\ref{fig:gramp}(b). The comparison includes the STA derived using the variational ansatz in Eq.~\eqref{eq:gtvar} (solid lines), the STA based on the self-similar ansatz in Eq.~\eqref{eq:gt} (dash-dotted lines), and a linear ramp (dashed lines). One can immediatly see that the STA designed using the variational ansatz achieves significantly higher fidelities compared to the other two approaches, especially for shorter timescales. The inset illustrates the energy difference between the final state obtained using the variational STA and linear ramps compared to the target state. For clarity, the self-similar STA, which exhibits a significantly higher energy difference, is not shown. Notably, even as the fidelity of linear ramps approaches unity, energy fluctuations persist, especially for slower ramps ( $t_f > 5$ ). In contrast, the variational STA demonstrates more stable dynamics, consistently achieving lower energy differences.

\section{\label{sec:engine} Application : Isentropic Engine}

The shortcuts we have developed enable precise control over all possible parameter variations in our system. An important application of STAs lies in enhancing the performance of quantum thermal machines, such as quantum heat engines. These shortcuts can be utilized for state preparation and to optimize the performance, significantly boosting output power by enabling each stroke to be executed quickly and without friction \cite{Campo2014, Beau_2016, Baris_2019, Hartmann2020, Fogarty_2021, Pedram_2023, Lewis2024}. Recent experimental advances have sparked increasing interest in interaction-driven engine cycles \cite{Li2018, Chen2019, Keller2020, MOMO2023, Watson2023}, with Refs.\cite{Koch2023, Simmons2023} successfully realizing these types of engines using ultracold atoms. Here, we apply our STA protocols to a unitary engine cycle similar to the one demonstrated in the recent experiment by Simmons et al.\cite{Simmons2023}.  In this work, a fully isentropic engine cycle was realized by replacing the conventional heating and cooling strokes with interaction-driven strokes. This engine pumps energy from the magnetic field to the optical trapping field via the trapped BEC, demonstrating enhanced efficiency and power compared to its non-degenerate gas counterpart.

We simulate the engine cycle using STA paths for all strokes and compare the results with those obtained from simple linear ramps. For this we assume that the BEC is initially trapped in an anisotropic potential with trap frequencies $\omega_l = (0.4, 0.5, 0.7)$. To avoid entering regimes of attractive interactions during fast interaction STAs, we consider a condensate firmly within the Thomas–Fermi (TF) regime \cite{Keller2020}, with an initial scattering length $a_l = 4000$.

\begin{figure}
    \centering
    \includegraphics[width=\linewidth]{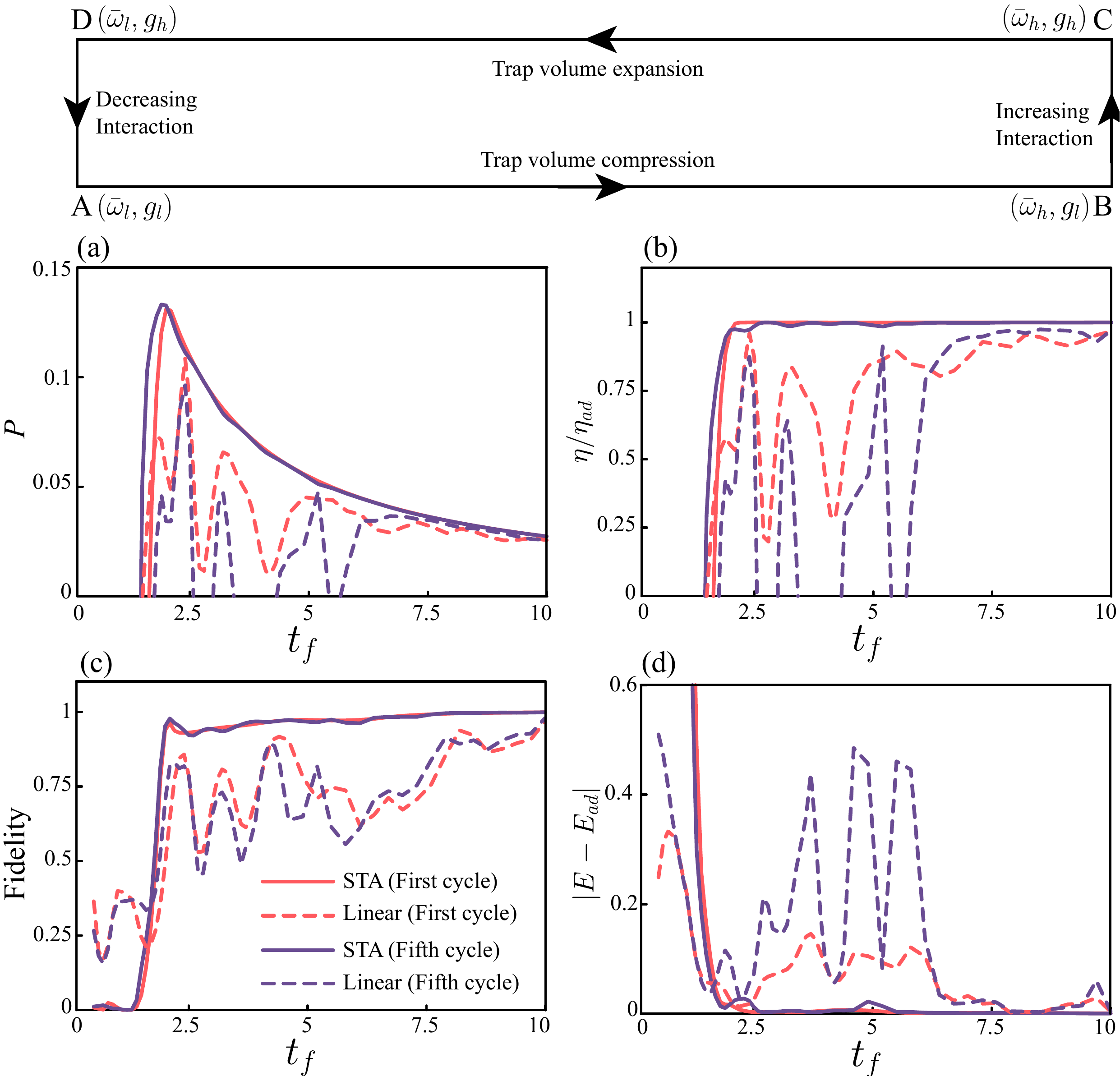}
    \caption{\textit{Top:} Schematic representation of the discussed isentropic engine cycle. The starting point is node A. The mean trap frequency $\bar{\omega}$ and interaction strength $g$ are indicated at all four nodes. Here, $\bar{\omega}_l = 0.53$ and $\bar{\omega}_h = 2\bar{\omega}_l$. For simplicity, the duration of each stroke is assumed to be equal. The interaction strengths are $g_l = 4000$ and $g_h = 6000$. (a) Comparison of engine power between an STA-driven cycle (solid lines) and a linearly driven cycle (dashed lines). The red and purple curves represent the power after the first and fifth cycles, respectively. (b) Efficiency of the driven cycle, scaled by the adiabatic efficiency.(c) Fidelity of the BEC at node A, determined by the overlap between the adiabatic state and the driven states after one and five full cycles. (d) Energy difference between the driven and adiabatic states at node A. The legends in (c) apply to all four panels.}
    \label{fig:cycle}
\end{figure}

The first stroke, $A \rightarrow B$, in the engine cycle (illustrated in Fig.~\ref{fig:cycle}), compresses the trap frequencies to double their initial mean value, $\omega_h = (1.4, 1, 0.8)$, while maintaining a constant interaction strength, resulting in work $W_{AB} = E_B - E_A$. The second stroke, $B \rightarrow C$, involves an interaction-driven process at fixed trap frequencies $\omega_h$, where the scattering length is increased to $a_h = 6000$, with an associated energy change $W_{BC} = E_C - E_B$. In the third stroke, $C \rightarrow D$, the trap frequencies are expanded from $\omega_h$ back to $\omega_l$ at a fixed scattering length $a_h$, extracting work $W_{CD} = E_D - E_C$. Finally, in the fourth stroke, $D \rightarrow A$, the interaction strength is returned to its initial value $a_l$, resulting in an energy change $W_{AD} = E_A - E_D$, completing one full cycle.

The engine performance is characterized by its efficiency, $\eta$, and power, $P$, defined as
\begin{align}
    \eta = -\dfrac{W_{AB}+W_{CD}}{W_{BC}},\\
    P = -\dfrac{W_{AB}+W_{CD}}{t_\text{cycle}},
\end{align}
where work is successfully output if $W_{AB} + W_{CD} \leq 0$. The total cycle time, $t_{\text{cycle}}$, is the sum of the durations of all four strokes. For simplicity, we assume equal duration for each stroke, $t_f$, giving $t_{\text{cycle}} = 4t_f$.

Figure \ref{fig:cycle} highlights the stark contrast between a linearly driven engine and an STA-driven engine cycle. In Fig.~\ref{fig:cycle}(a), the engine power is plotted as a function of ramp duration $t_f$. On short timescales, the STA-driven engine produces stable and high power, which remains consistent even after five consecutive cycles. In contrast, the power output from linear ramps fluctuates significantly with $t_f$ and progressively deteriorates with each subsequent cycle due to the accumulation of non-adiabatic excitations in the system. Similarly, the efficiency of the STA-driven engine, shown in Fig.~\ref{fig:cycle}(b), reaches its adiabatic value much earlier than the linear ramps, with only minor fluctuations observed when operating the engine over multiple cycles. 

The improved performance of the STA-driven engine can be attributed to its ability to closely approach the stationary adiabatic state within a given time. This is evident in Fig.~\ref{fig:cycle}(c), where the fidelity of STA-driven states approaches 1 after a full cycle and remains near unity even after five consecutive cycles, even for very short timescales. In contrast, linearly driven cycles exhibit oscillatory behavior, far from the maximum possible fidelity, clearly indicating that the stationary ground state has not been reached. These conclusions are further supported by Fig.~\ref{fig:cycle}(d), where we show the energy differences between the driven states and the adiabatic states at point A. The figure demonstrates the accumulation of unwanted excitations in the linearly driven system. Notably, the STA-driven engine performs exceptionally well, despite the approximate nature of the STAs and the absence of a dissipation mechanism to eliminate the small irreversible excitations they create.


\section{\label{sec:conclusion} Conclusion}

In summary, we have derived Ermakov-like equations for a BEC in general 3D anisotropic harmonic traps, with scaling parameters applicable across regimes ranging from non-interacting to the Thomas–Fermi limit. Following this, we employed inverse engineering protocols to design general STAs based on simultaneous changes in all three trap frequencies. These STAs are effective even for large structural transitions in the confining geometry; however, for larger interaction strengths, the minimum time required to achieve maximum fidelity increases.

For cases involving changes in interaction strength within a fixed trap geometry, the scaling approach proves effective for isotropic traps but becomes less reliable in anisotropic traps where the coupling between different spatial directions is strong. In such cases, we employed a variational approach to design alternative STAs, noting that these methods are only effective within the TF regime. We also highlighted the importance of using accurate anisotropic shortcuts, even for mildly anisotropic 3D traps. Approximate shortcuts based on mean trap frequencies, without accounting for anisotropy, were shown to perform worse than simple linear ramps.

As a significant application, we demonstrated that STAs can enhance the performance of BEC-based quantum engines by accelerating adiabatic strokes. In the specific case studied, the peak power output increased fivefold compared to the adiabatic driving limit at $t_f = 10$, while maintaining high efficiency. Furthermore, we showed that this enhanced performance persists over multiple cycles, primarily due to the STA protocols’ ability to consistently achieve states close to the stationary energy levels across repeated cycles.

An important extension of this work would be to consider the effects of temperature, which would enable simulations of more traditional Otto engine cycles. Such an approach would require modeling thermal BECs using c-field methods, such as the stochastic projected GPE, and could be employed to explore the optimization of BEC-enhanced engine cycles \cite{Myers2022}.

We thank Xi Chen for fruitful discussions. The work is supported by the Okinawa Institute of Science and Technology Graduate University (OIST). All numerical simulations  were performed on the high-performance computing cluster provided by the Scientific Computing and Data Analysis section at OIST. T.B. and T.F. acknowledge support from JST COI-NEXT Grant No. JPMJPF2221 and T.F. is also supported by JSPS KAKENHI Grant No. JP23K03290.

\appendix

\section{\label{app:3dermakov} Derivation of 3D Ermakov-like Equations}

We start with Eq.~\eqref{eq:hydrodyanmic}, which naturally decomposes into three equations corresponding to the three spatial directions. By substituting the velocity expression from the main text, $v_{\sigma \in (x, y, z)} = \dot{\sigma}_0 \sigma / \sigma_0$, into Eq.~\eqref{eq:hydrodyanmic} and performing the spatial integration, we obtain the following equations
\begin{align}
    -\dfrac{\ddot{x_0}x^2}{2x_0} = f(\textbf{r}, t) + \dfrac{\omega_x^2 x^2}{2} - f(\textbf{r}, t)|_{x=0},\\
    -\dfrac{\ddot{y_0}y^2}{2y_0} = f(\textbf{r}, t) + \dfrac{\omega_y^2 y^2}{2} - f(\textbf{r}, t)|_{y=0},\\
    -\dfrac{\ddot{z_0}z^2}{2z_0} = f(\textbf{r}, t) + \dfrac{\omega_z^2 z^2}{2} - f(\textbf{r}, t)|_{z=0},
\end{align}
where $f(\textbf{r}, t) = P(\textbf{r}, t) + gn(\textbf{r},t)$. Next, we introduce the re-scaled units, multiply both sides by $n_0(R, t)$, and perform another integration. This process yields Eqs.~\eqref{eq:trap3der1}–\eqref{eq:trap3der3} in the main text. The coefficients $A_{\sigma'}^\sigma/\omega_{\sigma}(0)^2$ and $B_\sigma/\omega_{\sigma}(0)^2$ characterize the different interaction regimes and should sum to one for each equation, in agreement with the one-dimensional case \cite{Huang2021, Modugno2018}. To confirm this, we analyze the system in its initial stationary state at time $t = 0$
\begin{align}
    \mu\Psi(R,t) = \left[-\dfrac{\nabla_R^2}{2}+V(R, t)+g(t)|\Psi(R, t)|^2\right]\Psi(R,t)\label{eq:stngpe}
\end{align}
and set $\Psi = \sqrt{n_0}$ at $x=0$. This gives an expression for $\mu$ as 
\begin{align}
    \mu = -P(R, t)|_{X=0} + \dfrac{1}{2}(\omega_x^2 X^2+\omega_y^2Y^2)+gn_0|_{X=0}
\end{align}
Substituting the expression for $\mu$ into Eq.~\eqref{eq:stngpe}, multiplying through by $n_0$, and performing the integration, we obtain the following result
\begin{align}
    A_x^x+A_y^x+A_z^x+B_x = \omega_x^2(0)
\end{align}
A similar approach for the other two directions can confirm that $A_x^y+A_y^y+A_z^y+B_y = \omega_y^2(0)$ and $A_x^z+A_y^z+A_z^z+B_z = \omega_z^2(0)$.

\bibliography{apssamp}

\begin{thebibliography}{66}%
\makeatletter
\providecommand \@ifxundefined [1]{%
 \@ifx{#1\undefined}
}%
\providecommand \@ifnum [1]{%
 \ifnum #1\expandafter \@firstoftwo
 \else \expandafter \@secondoftwo
 \fi
}%
\providecommand \@ifx [1]{%
 \ifx #1\expandafter \@firstoftwo
 \else \expandafter \@secondoftwo
 \fi
}%
\providecommand \natexlab [1]{#1}%
\providecommand \enquote  [1]{``#1''}%
\providecommand \bibnamefont  [1]{#1}%
\providecommand \bibfnamefont [1]{#1}%
\providecommand \citenamefont [1]{#1}%
\providecommand \href@noop [0]{\@secondoftwo}%
\providecommand \href [0]{\begingroup \@sanitize@url \@href}%
\providecommand \@href[1]{\@@startlink{#1}\@@href}%
\providecommand \@@href[1]{\endgroup#1\@@endlink}%
\providecommand \@sanitize@url [0]{\catcode `\\12\catcode `\$12\catcode `\&12\catcode `\#12\catcode `\^12\catcode `\_12\catcode `\%12\relax}%
\providecommand \@@startlink[1]{}%
\providecommand \@@endlink[0]{}%
\providecommand \url  [0]{\begingroup\@sanitize@url \@url }%
\providecommand \@url [1]{\endgroup\@href {#1}{\urlprefix }}%
\providecommand \urlprefix  [0]{URL }%
\providecommand \Eprint [0]{\href }%
\providecommand \doibase [0]{http://dx.doi.org/}%
\providecommand \selectlanguage [0]{\@gobble}%
\providecommand \bibinfo  [0]{\@secondoftwo}%
\providecommand \bibfield  [0]{\@secondoftwo}%
\providecommand \translation [1]{[#1]}%
\providecommand \BibitemOpen [0]{}%
\providecommand \bibitemStop [0]{}%
\providecommand \bibitemNoStop [0]{.\EOS\space}%
\providecommand \EOS [0]{\spacefactor3000\relax}%
\providecommand \BibitemShut  [1]{\csname bibitem#1\endcsname}%
\let\auto@bib@innerbib\@empty
\bibitem [{\citenamefont {Dalfovo}\ \emph {et~al.}(1999)\citenamefont {Dalfovo}, \citenamefont {Giorgini}, \citenamefont {Pitaevskii},\ and\ \citenamefont {Stringari}}]{Dalfovo1999}%
  \BibitemOpen
  \bibfield  {author} {\bibinfo {author} {\bibfnamefont {F.}~\bibnamefont {Dalfovo}}, \bibinfo {author} {\bibfnamefont {S.}~\bibnamefont {Giorgini}}, \bibinfo {author} {\bibfnamefont {L.~P.}\ \bibnamefont {Pitaevskii}}, \ and\ \bibinfo {author} {\bibfnamefont {S.}~\bibnamefont {Stringari}},\ }\href {\doibase 10.1103/RevModPhys.71.463} {\bibfield  {journal} {\bibinfo  {journal} {Rev. Mod. Phys.}\ }\textbf {\bibinfo {volume} {71}},\ \bibinfo {pages} {463} (\bibinfo {year} {1999})}\BibitemShut {NoStop}%
\bibitem [{\citenamefont {Chin}\ \emph {et~al.}(2010)\citenamefont {Chin}, \citenamefont {Grimm}, \citenamefont {Julienne},\ and\ \citenamefont {Tiesinga}}]{Chin2010}%
  \BibitemOpen
  \bibfield  {author} {\bibinfo {author} {\bibfnamefont {C.}~\bibnamefont {Chin}}, \bibinfo {author} {\bibfnamefont {R.}~\bibnamefont {Grimm}}, \bibinfo {author} {\bibfnamefont {P.}~\bibnamefont {Julienne}}, \ and\ \bibinfo {author} {\bibfnamefont {E.}~\bibnamefont {Tiesinga}},\ }\href {\doibase 10.1103/RevModPhys.82.1225} {\bibfield  {journal} {\bibinfo  {journal} {Rev. Mod. Phys.}\ }\textbf {\bibinfo {volume} {82}},\ \bibinfo {pages} {1225} (\bibinfo {year} {2010})}\BibitemShut {NoStop}%
\bibitem [{\citenamefont {Stringari}(1996)}]{Stringari1996}%
  \BibitemOpen
  \bibfield  {author} {\bibinfo {author} {\bibfnamefont {S.}~\bibnamefont {Stringari}},\ }\href {\doibase 10.1103/PhysRevLett.77.2360} {\bibfield  {journal} {\bibinfo  {journal} {Phys. Rev. Lett.}\ }\textbf {\bibinfo {volume} {77}},\ \bibinfo {pages} {2360} (\bibinfo {year} {1996})}\BibitemShut {NoStop}%
\bibitem [{\citenamefont {P\'erez-Garc\'{\i}a}\ \emph {et~al.}(1996)\citenamefont {P\'erez-Garc\'{\i}a}, \citenamefont {Michinel}, \citenamefont {Cirac}, \citenamefont {Lewenstein},\ and\ \citenamefont {Zoller}}]{Perez1996}%
  \BibitemOpen
  \bibfield  {author} {\bibinfo {author} {\bibfnamefont {V.~M.}\ \bibnamefont {P\'erez-Garc\'{\i}a}}, \bibinfo {author} {\bibfnamefont {H.}~\bibnamefont {Michinel}}, \bibinfo {author} {\bibfnamefont {J.~I.}\ \bibnamefont {Cirac}}, \bibinfo {author} {\bibfnamefont {M.}~\bibnamefont {Lewenstein}}, \ and\ \bibinfo {author} {\bibfnamefont {P.}~\bibnamefont {Zoller}},\ }\href {\doibase 10.1103/PhysRevLett.77.5320} {\bibfield  {journal} {\bibinfo  {journal} {Phys. Rev. Lett.}\ }\textbf {\bibinfo {volume} {77}},\ \bibinfo {pages} {5320} (\bibinfo {year} {1996})}\BibitemShut {NoStop}%
\bibitem [{\citenamefont {Andrews}\ \emph {et~al.}(1997)\citenamefont {Andrews}, \citenamefont {Kurn}, \citenamefont {Miesner}, \citenamefont {Durfee}, \citenamefont {Townsend}, \citenamefont {Inouye},\ and\ \citenamefont {Ketterle}}]{Andrews1997}%
  \BibitemOpen
  \bibfield  {author} {\bibinfo {author} {\bibfnamefont {M.~R.}\ \bibnamefont {Andrews}}, \bibinfo {author} {\bibfnamefont {D.~M.}\ \bibnamefont {Kurn}}, \bibinfo {author} {\bibfnamefont {H.-J.}\ \bibnamefont {Miesner}}, \bibinfo {author} {\bibfnamefont {D.~S.}\ \bibnamefont {Durfee}}, \bibinfo {author} {\bibfnamefont {C.~G.}\ \bibnamefont {Townsend}}, \bibinfo {author} {\bibfnamefont {S.}~\bibnamefont {Inouye}}, \ and\ \bibinfo {author} {\bibfnamefont {W.}~\bibnamefont {Ketterle}},\ }\href {\doibase 10.1103/PhysRevLett.79.553} {\bibfield  {journal} {\bibinfo  {journal} {Phys. Rev. Lett.}\ }\textbf {\bibinfo {volume} {79}},\ \bibinfo {pages} {553} (\bibinfo {year} {1997})}\BibitemShut {NoStop}%
\bibitem [{\citenamefont {Kagan}\ \emph {et~al.}(1997)\citenamefont {Kagan}, \citenamefont {Surkov},\ and\ \citenamefont {Shlyapnikov}}]{Kagan1997}%
  \BibitemOpen
  \bibfield  {author} {\bibinfo {author} {\bibfnamefont {Y.}~\bibnamefont {Kagan}}, \bibinfo {author} {\bibfnamefont {E.~L.}\ \bibnamefont {Surkov}}, \ and\ \bibinfo {author} {\bibfnamefont {G.~V.}\ \bibnamefont {Shlyapnikov}},\ }\href {\doibase 10.1103/PhysRevLett.79.2604} {\bibfield  {journal} {\bibinfo  {journal} {Phys. Rev. Lett.}\ }\textbf {\bibinfo {volume} {79}},\ \bibinfo {pages} {2604} (\bibinfo {year} {1997})}\BibitemShut {NoStop}%
\bibitem [{\citenamefont {Polkovnikov}\ \emph {et~al.}(2011)\citenamefont {Polkovnikov}, \citenamefont {Sengupta}, \citenamefont {Silva},\ and\ \citenamefont {Vengalattore}}]{Polkovnikov2011}%
  \BibitemOpen
  \bibfield  {author} {\bibinfo {author} {\bibfnamefont {A.}~\bibnamefont {Polkovnikov}}, \bibinfo {author} {\bibfnamefont {K.}~\bibnamefont {Sengupta}}, \bibinfo {author} {\bibfnamefont {A.}~\bibnamefont {Silva}}, \ and\ \bibinfo {author} {\bibfnamefont {M.}~\bibnamefont {Vengalattore}},\ }\href {\doibase 10.1103/RevModPhys.83.863} {\bibfield  {journal} {\bibinfo  {journal} {Rev. Mod. Phys.}\ }\textbf {\bibinfo {volume} {83}},\ \bibinfo {pages} {863} (\bibinfo {year} {2011})}\BibitemShut {NoStop}%
\bibitem [{\citenamefont {Cazalilla}\ and\ \citenamefont {Rigol}(2010)}]{Cazalilla2010}%
  \BibitemOpen
  \bibfield  {author} {\bibinfo {author} {\bibfnamefont {M.~A.}\ \bibnamefont {Cazalilla}}\ and\ \bibinfo {author} {\bibfnamefont {M.}~\bibnamefont {Rigol}},\ }\href {\doibase 10.1088/1367-2630/12/5/055006} {\bibfield  {journal} {\bibinfo  {journal} {New Journal of Physics}\ }\textbf {\bibinfo {volume} {12}},\ \bibinfo {pages} {055006} (\bibinfo {year} {2010})}\BibitemShut {NoStop}%
\bibitem [{\citenamefont {van Abeelen}\ and\ \citenamefont {Verhaar}(1999)}]{Abeelen1999}%
  \BibitemOpen
  \bibfield  {author} {\bibinfo {author} {\bibfnamefont {F.~A.}\ \bibnamefont {van Abeelen}}\ and\ \bibinfo {author} {\bibfnamefont {B.~J.}\ \bibnamefont {Verhaar}},\ }\href {\doibase 10.1103/PhysRevLett.83.1550} {\bibfield  {journal} {\bibinfo  {journal} {Phys. Rev. Lett.}\ }\textbf {\bibinfo {volume} {83}},\ \bibinfo {pages} {1550} (\bibinfo {year} {1999})}\BibitemShut {NoStop}%
\bibitem [{\citenamefont {Dieckmann}\ \emph {et~al.}(2002)\citenamefont {Dieckmann}, \citenamefont {Stan}, \citenamefont {Gupta}, \citenamefont {Hadzibabic}, \citenamefont {Schunck},\ and\ \citenamefont {Ketterle}}]{Ketterle2002}%
  \BibitemOpen
  \bibfield  {author} {\bibinfo {author} {\bibfnamefont {K.}~\bibnamefont {Dieckmann}}, \bibinfo {author} {\bibfnamefont {C.~A.}\ \bibnamefont {Stan}}, \bibinfo {author} {\bibfnamefont {S.}~\bibnamefont {Gupta}}, \bibinfo {author} {\bibfnamefont {Z.}~\bibnamefont {Hadzibabic}}, \bibinfo {author} {\bibfnamefont {C.~H.}\ \bibnamefont {Schunck}}, \ and\ \bibinfo {author} {\bibfnamefont {W.}~\bibnamefont {Ketterle}},\ }\href {\doibase 10.1103/PhysRevLett.89.203201} {\bibfield  {journal} {\bibinfo  {journal} {Phys. Rev. Lett.}\ }\textbf {\bibinfo {volume} {89}},\ \bibinfo {pages} {203201} (\bibinfo {year} {2002})}\BibitemShut {NoStop}%
\bibitem [{\citenamefont {Regal}\ \emph {et~al.}(2004)\citenamefont {Regal}, \citenamefont {Greiner},\ and\ \citenamefont {Jin}}]{Regal2004}%
  \BibitemOpen
  \bibfield  {author} {\bibinfo {author} {\bibfnamefont {C.~A.}\ \bibnamefont {Regal}}, \bibinfo {author} {\bibfnamefont {M.}~\bibnamefont {Greiner}}, \ and\ \bibinfo {author} {\bibfnamefont {D.~S.}\ \bibnamefont {Jin}},\ }\href {\doibase 10.1103/PhysRevLett.92.083201} {\bibfield  {journal} {\bibinfo  {journal} {Phys. Rev. Lett.}\ }\textbf {\bibinfo {volume} {92}},\ \bibinfo {pages} {083201} (\bibinfo {year} {2004})}\BibitemShut {NoStop}%
\bibitem [{\citenamefont {Carvalho}\ \emph {et~al.}(2004)\citenamefont {Carvalho}, \citenamefont {Mintert},\ and\ \citenamefont {Buchleitner}}]{Carvalho2004}%
  \BibitemOpen
  \bibfield  {author} {\bibinfo {author} {\bibfnamefont {A.~R.~R.}\ \bibnamefont {Carvalho}}, \bibinfo {author} {\bibfnamefont {F.}~\bibnamefont {Mintert}}, \ and\ \bibinfo {author} {\bibfnamefont {A.}~\bibnamefont {Buchleitner}},\ }\href {\doibase 10.1103/PhysRevLett.93.230501} {\bibfield  {journal} {\bibinfo  {journal} {Phys. Rev. Lett.}\ }\textbf {\bibinfo {volume} {93}},\ \bibinfo {pages} {230501} (\bibinfo {year} {2004})}\BibitemShut {NoStop}%
\bibitem [{\citenamefont {Gu\'ery-Odelin}\ \emph {et~al.}(2019)\citenamefont {Gu\'ery-Odelin}, \citenamefont {Ruschhaupt}, \citenamefont {Kiely}, \citenamefont {Torrontegui}, \citenamefont {Mart\'{\i}nez-Garaot},\ and\ \citenamefont {Muga}}]{Odelin2019}%
  \BibitemOpen
  \bibfield  {author} {\bibinfo {author} {\bibfnamefont {D.}~\bibnamefont {Gu\'ery-Odelin}}, \bibinfo {author} {\bibfnamefont {A.}~\bibnamefont {Ruschhaupt}}, \bibinfo {author} {\bibfnamefont {A.}~\bibnamefont {Kiely}}, \bibinfo {author} {\bibfnamefont {E.}~\bibnamefont {Torrontegui}}, \bibinfo {author} {\bibfnamefont {S.}~\bibnamefont {Mart\'{\i}nez-Garaot}}, \ and\ \bibinfo {author} {\bibfnamefont {J.~G.}\ \bibnamefont {Muga}},\ }\href {\doibase 10.1103/RevModPhys.91.045001} {\bibfield  {journal} {\bibinfo  {journal} {Rev. Mod. Phys.}\ }\textbf {\bibinfo {volume} {91}},\ \bibinfo {pages} {045001} (\bibinfo {year} {2019})}\BibitemShut {NoStop}%
\bibitem [{\citenamefont {Torrontegui}\ \emph {et~al.}(2014)\citenamefont {Torrontegui}, \citenamefont {Mart\'{\i}nez-Garaot},\ and\ \citenamefont {Muga}}]{Torrontegui2014}%
  \BibitemOpen
  \bibfield  {author} {\bibinfo {author} {\bibfnamefont {E.}~\bibnamefont {Torrontegui}}, \bibinfo {author} {\bibfnamefont {S.}~\bibnamefont {Mart\'{\i}nez-Garaot}}, \ and\ \bibinfo {author} {\bibfnamefont {J.~G.}\ \bibnamefont {Muga}},\ }\href {\doibase 10.1103/PhysRevA.89.043408} {\bibfield  {journal} {\bibinfo  {journal} {Phys. Rev. A}\ }\textbf {\bibinfo {volume} {89}},\ \bibinfo {pages} {043408} (\bibinfo {year} {2014})}\BibitemShut {NoStop}%
\bibitem [{\citenamefont {Vitanov}\ and\ \citenamefont {Shore}(2015)}]{Vitanov2015}%
  \BibitemOpen
  \bibfield  {author} {\bibinfo {author} {\bibfnamefont {N.~V.}\ \bibnamefont {Vitanov}}\ and\ \bibinfo {author} {\bibfnamefont {B.~W.}\ \bibnamefont {Shore}},\ }\href {\doibase 10.1088/0953-4075/48/17/174008} {\bibfield  {journal} {\bibinfo  {journal} {Journal of Physics B: Atomic, Molecular and Optical Physics}\ }\textbf {\bibinfo {volume} {48}},\ \bibinfo {pages} {174008} (\bibinfo {year} {2015})}\BibitemShut {NoStop}%
\bibitem [{\citenamefont {Zhang}\ \emph {et~al.}(2017)\citenamefont {Zhang}, \citenamefont {Chen},\ and\ \citenamefont {Gu{\'e}ry-Odelin}}]{Zhang2017}%
  \BibitemOpen
  \bibfield  {author} {\bibinfo {author} {\bibfnamefont {Q.}~\bibnamefont {Zhang}}, \bibinfo {author} {\bibfnamefont {X.}~\bibnamefont {Chen}}, \ and\ \bibinfo {author} {\bibfnamefont {D.}~\bibnamefont {Gu{\'e}ry-Odelin}},\ }\href {\doibase 10.1038/s41598-017-16146-2} {\bibfield  {journal} {\bibinfo  {journal} {Scientific Reports}\ }\textbf {\bibinfo {volume} {7}},\ \bibinfo {pages} {15814} (\bibinfo {year} {2017})}\BibitemShut {NoStop}%
\bibitem [{\citenamefont {Impens}\ and\ \citenamefont {Gu\'ery-Odelin}(2017)}]{Impens2017}%
  \BibitemOpen
  \bibfield  {author} {\bibinfo {author} {\bibfnamefont {F.}~\bibnamefont {Impens}}\ and\ \bibinfo {author} {\bibfnamefont {D.}~\bibnamefont {Gu\'ery-Odelin}},\ }\href {\doibase 10.1103/PhysRevA.96.043609} {\bibfield  {journal} {\bibinfo  {journal} {Phys. Rev. A}\ }\textbf {\bibinfo {volume} {96}},\ \bibinfo {pages} {043609} (\bibinfo {year} {2017})}\BibitemShut {NoStop}%
\bibitem [{\citenamefont {Berry}(2009)}]{Berry2009}%
  \BibitemOpen
  \bibfield  {author} {\bibinfo {author} {\bibfnamefont {M.~V.}\ \bibnamefont {Berry}},\ }\href {\doibase 10.1088/1751-8113/42/36/365303} {\bibfield  {journal} {\bibinfo  {journal} {Journal of Physics A: Mathematical and Theoretical}\ }\textbf {\bibinfo {volume} {42}},\ \bibinfo {pages} {365303} (\bibinfo {year} {2009})}\BibitemShut {NoStop}%
\bibitem [{\citenamefont {del Campo}(2013)}]{Campo2013}%
  \BibitemOpen
  \bibfield  {author} {\bibinfo {author} {\bibfnamefont {A.}~\bibnamefont {del Campo}},\ }\href {\doibase 10.1103/PhysRevLett.111.100502} {\bibfield  {journal} {\bibinfo  {journal} {Phys. Rev. Lett.}\ }\textbf {\bibinfo {volume} {111}},\ \bibinfo {pages} {100502} (\bibinfo {year} {2013})}\BibitemShut {NoStop}%
\bibitem [{\citenamefont {Chen}\ \emph {et~al.}(2010)\citenamefont {Chen}, \citenamefont {Ruschhaupt}, \citenamefont {Schmidt}, \citenamefont {del Campo}, \citenamefont {Gu\'ery-Odelin},\ and\ \citenamefont {Muga}}]{Chen2010}%
  \BibitemOpen
  \bibfield  {author} {\bibinfo {author} {\bibfnamefont {X.}~\bibnamefont {Chen}}, \bibinfo {author} {\bibfnamefont {A.}~\bibnamefont {Ruschhaupt}}, \bibinfo {author} {\bibfnamefont {S.}~\bibnamefont {Schmidt}}, \bibinfo {author} {\bibfnamefont {A.}~\bibnamefont {del Campo}}, \bibinfo {author} {\bibfnamefont {D.}~\bibnamefont {Gu\'ery-Odelin}}, \ and\ \bibinfo {author} {\bibfnamefont {J.~G.}\ \bibnamefont {Muga}},\ }\href {\doibase 10.1103/PhysRevLett.104.063002} {\bibfield  {journal} {\bibinfo  {journal} {Phys. Rev. Lett.}\ }\textbf {\bibinfo {volume} {104}},\ \bibinfo {pages} {063002} (\bibinfo {year} {2010})}\BibitemShut {NoStop}%
\bibitem [{\citenamefont {Li}\ \emph {et~al.}(2016)\citenamefont {Li}, \citenamefont {Sun},\ and\ \citenamefont {Chen}}]{Li2016}%
  \BibitemOpen
  \bibfield  {author} {\bibinfo {author} {\bibfnamefont {J.}~\bibnamefont {Li}}, \bibinfo {author} {\bibfnamefont {K.}~\bibnamefont {Sun}}, \ and\ \bibinfo {author} {\bibfnamefont {X.}~\bibnamefont {Chen}},\ }\href {\doibase 10.1038/srep38258} {\bibfield  {journal} {\bibinfo  {journal} {Scientific Reports}\ }\textbf {\bibinfo {volume} {6}},\ \bibinfo {pages} {38258} (\bibinfo {year} {2016})}\BibitemShut {NoStop}%
\bibitem [{\citenamefont {Li}\ \emph {et~al.}(2018)\citenamefont {Li}, \citenamefont {Fogarty}, \citenamefont {Campbell}, \citenamefont {Chen},\ and\ \citenamefont {Busch}}]{Li2018}%
  \BibitemOpen
  \bibfield  {author} {\bibinfo {author} {\bibfnamefont {J.}~\bibnamefont {Li}}, \bibinfo {author} {\bibfnamefont {T.}~\bibnamefont {Fogarty}}, \bibinfo {author} {\bibfnamefont {S.}~\bibnamefont {Campbell}}, \bibinfo {author} {\bibfnamefont {X.}~\bibnamefont {Chen}}, \ and\ \bibinfo {author} {\bibfnamefont {T.}~\bibnamefont {Busch}},\ }\href {\doibase 10.1088/1367-2630/aa9cd8} {\bibfield  {journal} {\bibinfo  {journal} {New Journal of Physics}\ }\textbf {\bibinfo {volume} {20}},\ \bibinfo {pages} {015005} (\bibinfo {year} {2018})}\BibitemShut {NoStop}%
\bibitem [{\citenamefont {Masuda}\ and\ \citenamefont {Nakamura}(2008)}]{Masuda2008}%
  \BibitemOpen
  \bibfield  {author} {\bibinfo {author} {\bibfnamefont {S.}~\bibnamefont {Masuda}}\ and\ \bibinfo {author} {\bibfnamefont {K.}~\bibnamefont {Nakamura}},\ }\href {\doibase 10.1103/PhysRevA.78.062108} {\bibfield  {journal} {\bibinfo  {journal} {Phys. Rev. A}\ }\textbf {\bibinfo {volume} {78}},\ \bibinfo {pages} {062108} (\bibinfo {year} {2008})}\BibitemShut {NoStop}%
\bibitem [{\citenamefont {Schaff}\ \emph {et~al.}(2010)\citenamefont {Schaff}, \citenamefont {Song}, \citenamefont {Vignolo},\ and\ \citenamefont {Labeyrie}}]{Schaff2010}%
  \BibitemOpen
  \bibfield  {author} {\bibinfo {author} {\bibfnamefont {J.-F.}\ \bibnamefont {Schaff}}, \bibinfo {author} {\bibfnamefont {X.-L.}\ \bibnamefont {Song}}, \bibinfo {author} {\bibfnamefont {P.}~\bibnamefont {Vignolo}}, \ and\ \bibinfo {author} {\bibfnamefont {G.}~\bibnamefont {Labeyrie}},\ }\href {\doibase 10.1103/PhysRevA.82.033430} {\bibfield  {journal} {\bibinfo  {journal} {Phys. Rev. A}\ }\textbf {\bibinfo {volume} {82}},\ \bibinfo {pages} {033430} (\bibinfo {year} {2010})}\BibitemShut {NoStop}%
\bibitem [{\citenamefont {Schaff}\ \emph {et~al.}(2011)\citenamefont {Schaff}, \citenamefont {Song}, \citenamefont {Capuzzi}, \citenamefont {Vignolo},\ and\ \citenamefont {Labeyrie}}]{Schaff2011}%
  \BibitemOpen
  \bibfield  {author} {\bibinfo {author} {\bibfnamefont {J.-F.}\ \bibnamefont {Schaff}}, \bibinfo {author} {\bibfnamefont {X.-L.}\ \bibnamefont {Song}}, \bibinfo {author} {\bibfnamefont {P.}~\bibnamefont {Capuzzi}}, \bibinfo {author} {\bibfnamefont {P.}~\bibnamefont {Vignolo}}, \ and\ \bibinfo {author} {\bibfnamefont {G.}~\bibnamefont {Labeyrie}},\ }\href {\doibase 10.1209/0295-5075/93/23001} {\bibfield  {journal} {\bibinfo  {journal} {Europhysics Letters}\ }\textbf {\bibinfo {volume} {93}},\ \bibinfo {pages} {23001} (\bibinfo {year} {2011})}\BibitemShut {NoStop}%
\bibitem [{\citenamefont {Bowler}\ \emph {et~al.}(2012)\citenamefont {Bowler}, \citenamefont {Gaebler}, \citenamefont {Lin}, \citenamefont {Tan}, \citenamefont {Hanneke}, \citenamefont {Jost}, \citenamefont {Home}, \citenamefont {Leibfried},\ and\ \citenamefont {Wineland}}]{Bowler2012}%
  \BibitemOpen
  \bibfield  {author} {\bibinfo {author} {\bibfnamefont {R.}~\bibnamefont {Bowler}}, \bibinfo {author} {\bibfnamefont {J.}~\bibnamefont {Gaebler}}, \bibinfo {author} {\bibfnamefont {Y.}~\bibnamefont {Lin}}, \bibinfo {author} {\bibfnamefont {T.~R.}\ \bibnamefont {Tan}}, \bibinfo {author} {\bibfnamefont {D.}~\bibnamefont {Hanneke}}, \bibinfo {author} {\bibfnamefont {J.~D.}\ \bibnamefont {Jost}}, \bibinfo {author} {\bibfnamefont {J.~P.}\ \bibnamefont {Home}}, \bibinfo {author} {\bibfnamefont {D.}~\bibnamefont {Leibfried}}, \ and\ \bibinfo {author} {\bibfnamefont {D.~J.}\ \bibnamefont {Wineland}},\ }\href {\doibase 10.1103/PhysRevLett.109.080502} {\bibfield  {journal} {\bibinfo  {journal} {Phys. Rev. Lett.}\ }\textbf {\bibinfo {volume} {109}},\ \bibinfo {pages} {080502} (\bibinfo {year} {2012})}\BibitemShut {NoStop}%
\bibitem [{\citenamefont {Rohringer}\ \emph {et~al.}(2015)\citenamefont {Rohringer}, \citenamefont {Fischer}, \citenamefont {Steiner}, \citenamefont {Mazets}, \citenamefont {Schmiedmayer},\ and\ \citenamefont {Trupke}}]{Rohringer2015}%
  \BibitemOpen
  \bibfield  {author} {\bibinfo {author} {\bibfnamefont {W.}~\bibnamefont {Rohringer}}, \bibinfo {author} {\bibfnamefont {D.}~\bibnamefont {Fischer}}, \bibinfo {author} {\bibfnamefont {F.}~\bibnamefont {Steiner}}, \bibinfo {author} {\bibfnamefont {I.~E.}\ \bibnamefont {Mazets}}, \bibinfo {author} {\bibfnamefont {J.}~\bibnamefont {Schmiedmayer}}, \ and\ \bibinfo {author} {\bibfnamefont {M.}~\bibnamefont {Trupke}},\ }\href {\doibase 10.1038/srep09820} {\bibfield  {journal} {\bibinfo  {journal} {Scientific Reports}\ }\textbf {\bibinfo {volume} {5}},\ \bibinfo {pages} {9820} (\bibinfo {year} {2015})}\BibitemShut {NoStop}%
\bibitem [{\citenamefont {Deng}\ \emph {et~al.}(2018)\citenamefont {Deng}, \citenamefont {Diao}, \citenamefont {Yu}, \citenamefont {del Campo},\ and\ \citenamefont {Wu}}]{Deng2018}%
  \BibitemOpen
  \bibfield  {author} {\bibinfo {author} {\bibfnamefont {S.}~\bibnamefont {Deng}}, \bibinfo {author} {\bibfnamefont {P.}~\bibnamefont {Diao}}, \bibinfo {author} {\bibfnamefont {Q.}~\bibnamefont {Yu}}, \bibinfo {author} {\bibfnamefont {A.}~\bibnamefont {del Campo}}, \ and\ \bibinfo {author} {\bibfnamefont {H.}~\bibnamefont {Wu}},\ }\href {\doibase 10.1103/PhysRevA.97.013628} {\bibfield  {journal} {\bibinfo  {journal} {Phys. Rev. A}\ }\textbf {\bibinfo {volume} {97}},\ \bibinfo {pages} {013628} (\bibinfo {year} {2018})}\BibitemShut {NoStop}%
\bibitem [{\citenamefont {del Campo}\ and\ \citenamefont {Boshier}(2012)}]{Campo2012}%
  \BibitemOpen
  \bibfield  {author} {\bibinfo {author} {\bibfnamefont {A.}~\bibnamefont {del Campo}}\ and\ \bibinfo {author} {\bibfnamefont {M.~G.}\ \bibnamefont {Boshier}},\ }\href {\doibase 10.1038/srep00648} {\bibfield  {journal} {\bibinfo  {journal} {Scientific Reports}\ }\textbf {\bibinfo {volume} {2}},\ \bibinfo {pages} {648} (\bibinfo {year} {2012})}\BibitemShut {NoStop}%
\bibitem [{\citenamefont {Beau}\ \emph {et~al.}(2016)\citenamefont {Beau}, \citenamefont {Jaramillo},\ and\ \citenamefont {Del~Campo}}]{Beau_2016}%
  \BibitemOpen
  \bibfield  {author} {\bibinfo {author} {\bibfnamefont {M.}~\bibnamefont {Beau}}, \bibinfo {author} {\bibfnamefont {J.}~\bibnamefont {Jaramillo}}, \ and\ \bibinfo {author} {\bibfnamefont {A.}~\bibnamefont {Del~Campo}},\ }\href {\doibase 10.3390/e18050168} {\bibfield  {journal} {\bibinfo  {journal} {Entropy}\ }\textbf {\bibinfo {volume} {18}} (\bibinfo {year} {2016}),\ 10.3390/e18050168}\BibitemShut {NoStop}%
\bibitem [{\citenamefont {Sels}\ and\ \citenamefont {Polkovnikov}(2017)}]{Sels2017}%
  \BibitemOpen
  \bibfield  {author} {\bibinfo {author} {\bibfnamefont {D.}~\bibnamefont {Sels}}\ and\ \bibinfo {author} {\bibfnamefont {A.}~\bibnamefont {Polkovnikov}},\ }\href {\doibase 10.1073/pnas.1619826114} {\bibfield  {journal} {\bibinfo  {journal} {Proceedings of the National Academy of Sciences}\ }\textbf {\bibinfo {volume} {114}},\ \bibinfo {pages} {E3909} (\bibinfo {year} {2017})},\ \Eprint {http://arxiv.org/abs/https://www.pnas.org/doi/pdf/10.1073/pnas.1619826114} {https://www.pnas.org/doi/pdf/10.1073/pnas.1619826114} \BibitemShut {NoStop}%
\bibitem [{\citenamefont {Xu}\ \emph {et~al.}(2020)\citenamefont {Xu}, \citenamefont {Li}, \citenamefont {Busch}, \citenamefont {Chen},\ and\ \citenamefont {Fogarty}}]{Skycow2020}%
  \BibitemOpen
  \bibfield  {author} {\bibinfo {author} {\bibfnamefont {T.-N.}\ \bibnamefont {Xu}}, \bibinfo {author} {\bibfnamefont {J.}~\bibnamefont {Li}}, \bibinfo {author} {\bibfnamefont {T.}~\bibnamefont {Busch}}, \bibinfo {author} {\bibfnamefont {X.}~\bibnamefont {Chen}}, \ and\ \bibinfo {author} {\bibfnamefont {T.}~\bibnamefont {Fogarty}},\ }\href {\doibase 10.1103/PhysRevResearch.2.023125} {\bibfield  {journal} {\bibinfo  {journal} {Phys. Rev. Res.}\ }\textbf {\bibinfo {volume} {2}},\ \bibinfo {pages} {023125} (\bibinfo {year} {2020})}\BibitemShut {NoStop}%
\bibitem [{\citenamefont {Fogarty}\ and\ \citenamefont {Busch}(2020)}]{Fogarty_2021}%
  \BibitemOpen
  \bibfield  {author} {\bibinfo {author} {\bibfnamefont {T.}~\bibnamefont {Fogarty}}\ and\ \bibinfo {author} {\bibfnamefont {T.}~\bibnamefont {Busch}},\ }\href {\doibase 10.1088/2058-9565/abbc63} {\bibfield  {journal} {\bibinfo  {journal} {Quantum Science and Technology}\ }\textbf {\bibinfo {volume} {6}},\ \bibinfo {pages} {015003} (\bibinfo {year} {2020})}\BibitemShut {NoStop}%
\bibitem [{\citenamefont {\ifmmode \check{C}\else \v{C}\fi{}epait\ifmmode~\dot{e}\else \.{e}\fi{}}\ \emph {et~al.}(2023)\citenamefont {\ifmmode \check{C}\else \v{C}\fi{}epait\ifmmode~\dot{e}\else \.{e}\fi{}}, \citenamefont {Polkovnikov}, \citenamefont {Daley},\ and\ \citenamefont {Duncan}}]{COLD_2023}%
  \BibitemOpen
  \bibfield  {author} {\bibinfo {author} {\bibfnamefont {I.}~\bibnamefont {\ifmmode \check{C}\else \v{C}\fi{}epait\ifmmode~\dot{e}\else \.{e}\fi{}}}, \bibinfo {author} {\bibfnamefont {A.}~\bibnamefont {Polkovnikov}}, \bibinfo {author} {\bibfnamefont {A.~J.}\ \bibnamefont {Daley}}, \ and\ \bibinfo {author} {\bibfnamefont {C.~W.}\ \bibnamefont {Duncan}},\ }\href {\doibase 10.1103/PRXQuantum.4.010312} {\bibfield  {journal} {\bibinfo  {journal} {PRX Quantum}\ }\textbf {\bibinfo {volume} {4}},\ \bibinfo {pages} {010312} (\bibinfo {year} {2023})}\BibitemShut {NoStop}%
\bibitem [{\citenamefont {Hasan}\ \emph {et~al.}(2024)\citenamefont {Hasan}, \citenamefont {Fogarty}, \citenamefont {Li}, \citenamefont {Ruschhaupt},\ and\ \citenamefont {Busch}}]{Hasan2024}%
  \BibitemOpen
  \bibfield  {author} {\bibinfo {author} {\bibfnamefont {M.~S.}\ \bibnamefont {Hasan}}, \bibinfo {author} {\bibfnamefont {T.}~\bibnamefont {Fogarty}}, \bibinfo {author} {\bibfnamefont {J.}~\bibnamefont {Li}}, \bibinfo {author} {\bibfnamefont {A.}~\bibnamefont {Ruschhaupt}}, \ and\ \bibinfo {author} {\bibfnamefont {T.}~\bibnamefont {Busch}},\ }\href {\doibase 10.1103/PhysRevResearch.6.023114} {\bibfield  {journal} {\bibinfo  {journal} {Phys. Rev. Res.}\ }\textbf {\bibinfo {volume} {6}},\ \bibinfo {pages} {023114} (\bibinfo {year} {2024})}\BibitemShut {NoStop}%
\bibitem [{\citenamefont {Morawetz}\ and\ \citenamefont {Polkovnikov}(2024)}]{Morawetz_2024}%
  \BibitemOpen
  \bibfield  {author} {\bibinfo {author} {\bibfnamefont {S.}~\bibnamefont {Morawetz}}\ and\ \bibinfo {author} {\bibfnamefont {A.}~\bibnamefont {Polkovnikov}},\ }\href {\doibase 10.1103/PhysRevB.110.024304} {\bibfield  {journal} {\bibinfo  {journal} {Phys. Rev. B}\ }\textbf {\bibinfo {volume} {110}},\ \bibinfo {pages} {024304} (\bibinfo {year} {2024})}\BibitemShut {NoStop}%
\bibitem [{\citenamefont {Deffner}\ \emph {et~al.}(2014)\citenamefont {Deffner}, \citenamefont {Jarzynski},\ and\ \citenamefont {del Campo}}]{Deffner2014}%
  \BibitemOpen
  \bibfield  {author} {\bibinfo {author} {\bibfnamefont {S.}~\bibnamefont {Deffner}}, \bibinfo {author} {\bibfnamefont {C.}~\bibnamefont {Jarzynski}}, \ and\ \bibinfo {author} {\bibfnamefont {A.}~\bibnamefont {del Campo}},\ }\href {\doibase 10.1103/PhysRevX.4.021013} {\bibfield  {journal} {\bibinfo  {journal} {Phys. Rev. X}\ }\textbf {\bibinfo {volume} {4}},\ \bibinfo {pages} {021013} (\bibinfo {year} {2014})}\BibitemShut {NoStop}%
\bibitem [{\citenamefont {Muga}\ \emph {et~al.}(2009)\citenamefont {Muga}, \citenamefont {Chen}, \citenamefont {Ruschhaupt},\ and\ \citenamefont {Guéry-Odelin}}]{Muga2009}%
  \BibitemOpen
  \bibfield  {author} {\bibinfo {author} {\bibfnamefont {J.~G.}\ \bibnamefont {Muga}}, \bibinfo {author} {\bibfnamefont {X.}~\bibnamefont {Chen}}, \bibinfo {author} {\bibfnamefont {A.}~\bibnamefont {Ruschhaupt}}, \ and\ \bibinfo {author} {\bibfnamefont {D.}~\bibnamefont {Guéry-Odelin}},\ }\href {\doibase 10.1088/0953-4075/42/24/241001} {\bibfield  {journal} {\bibinfo  {journal} {Journal of Physics B: Atomic, Molecular and Optical Physics}\ }\textbf {\bibinfo {volume} {42}},\ \bibinfo {pages} {241001} (\bibinfo {year} {2009})}\BibitemShut {NoStop}%
\bibitem [{\citenamefont {Gu\'ery-Odelin}(2002)}]{Odelin2002}%
  \BibitemOpen
  \bibfield  {author} {\bibinfo {author} {\bibfnamefont {D.}~\bibnamefont {Gu\'ery-Odelin}},\ }\href {\doibase 10.1103/PhysRevA.66.033613} {\bibfield  {journal} {\bibinfo  {journal} {Phys. Rev. A}\ }\textbf {\bibinfo {volume} {66}},\ \bibinfo {pages} {033613} (\bibinfo {year} {2002})}\BibitemShut {NoStop}%
\bibitem [{\citenamefont {Menotti}\ \emph {et~al.}(2002)\citenamefont {Menotti}, \citenamefont {Pedri},\ and\ \citenamefont {Stringari}}]{Menotti2002}%
  \BibitemOpen
  \bibfield  {author} {\bibinfo {author} {\bibfnamefont {C.}~\bibnamefont {Menotti}}, \bibinfo {author} {\bibfnamefont {P.}~\bibnamefont {Pedri}}, \ and\ \bibinfo {author} {\bibfnamefont {S.}~\bibnamefont {Stringari}},\ }\href {\doibase 10.1103/PhysRevLett.89.250402} {\bibfield  {journal} {\bibinfo  {journal} {Phys. Rev. Lett.}\ }\textbf {\bibinfo {volume} {89}},\ \bibinfo {pages} {250402} (\bibinfo {year} {2002})}\BibitemShut {NoStop}%
\bibitem [{\citenamefont {Hu}\ \emph {et~al.}(2003)\citenamefont {Hu}, \citenamefont {Liu},\ and\ \citenamefont {Modugno}}]{Hu2003}%
  \BibitemOpen
  \bibfield  {author} {\bibinfo {author} {\bibfnamefont {H.}~\bibnamefont {Hu}}, \bibinfo {author} {\bibfnamefont {X.-J.}\ \bibnamefont {Liu}}, \ and\ \bibinfo {author} {\bibfnamefont {M.}~\bibnamefont {Modugno}},\ }\href {\doibase 10.1103/PhysRevA.67.063614} {\bibfield  {journal} {\bibinfo  {journal} {Phys. Rev. A}\ }\textbf {\bibinfo {volume} {67}},\ \bibinfo {pages} {063614} (\bibinfo {year} {2003})}\BibitemShut {NoStop}%
\bibitem [{\citenamefont {Modugno}\ \emph {et~al.}(2018)\citenamefont {Modugno}, \citenamefont {Pagnini},\ and\ \citenamefont {Valle-Basagoiti}}]{Modugno2018}%
  \BibitemOpen
  \bibfield  {author} {\bibinfo {author} {\bibfnamefont {M.}~\bibnamefont {Modugno}}, \bibinfo {author} {\bibfnamefont {G.}~\bibnamefont {Pagnini}}, \ and\ \bibinfo {author} {\bibfnamefont {M.~A.}\ \bibnamefont {Valle-Basagoiti}},\ }\href {\doibase 10.1103/PhysRevA.97.043604} {\bibfield  {journal} {\bibinfo  {journal} {Phys. Rev. A}\ }\textbf {\bibinfo {volume} {97}},\ \bibinfo {pages} {043604} (\bibinfo {year} {2018})}\BibitemShut {NoStop}%
\bibitem [{\citenamefont {Viedma}\ and\ \citenamefont {Modugno}(2020)}]{Viedma2020}%
  \BibitemOpen
  \bibfield  {author} {\bibinfo {author} {\bibfnamefont {D.}~\bibnamefont {Viedma}}\ and\ \bibinfo {author} {\bibfnamefont {M.}~\bibnamefont {Modugno}},\ }\href {\doibase 10.1103/PhysRevResearch.2.033478} {\bibfield  {journal} {\bibinfo  {journal} {Phys. Rev. Res.}\ }\textbf {\bibinfo {volume} {2}},\ \bibinfo {pages} {033478} (\bibinfo {year} {2020})}\BibitemShut {NoStop}%
\bibitem [{\citenamefont {del Campo}\ \emph {et~al.}(2014{\natexlab{a}})\citenamefont {del Campo}, \citenamefont {Goold},\ and\ \citenamefont {Paternostro}}]{Campo2014-cf}%
  \BibitemOpen
  \bibfield  {author} {\bibinfo {author} {\bibfnamefont {A.}~\bibnamefont {del Campo}}, \bibinfo {author} {\bibfnamefont {J.}~\bibnamefont {Goold}}, \ and\ \bibinfo {author} {\bibfnamefont {M.}~\bibnamefont {Paternostro}},\ }\href@noop {} {\bibfield  {journal} {\bibinfo  {journal} {Scientific Reports}\ }\textbf {\bibinfo {volume} {4}},\ \bibinfo {pages} {6208} (\bibinfo {year} {2014}{\natexlab{a}})}\BibitemShut {NoStop}%
\bibitem [{\citenamefont {del Campo}(2011)}]{Campo2016}%
  \BibitemOpen
  \bibfield  {author} {\bibinfo {author} {\bibfnamefont {A.}~\bibnamefont {del Campo}},\ }\href {\doibase 10.1103/PhysRevA.84.031606} {\bibfield  {journal} {\bibinfo  {journal} {Phys. Rev. A}\ }\textbf {\bibinfo {volume} {84}},\ \bibinfo {pages} {031606} (\bibinfo {year} {2011})}\BibitemShut {NoStop}%
\bibitem [{\citenamefont {Myers}\ \emph {et~al.}(2022)\citenamefont {Myers}, \citenamefont {Peña}, \citenamefont {Negrete}, \citenamefont {Vargas}, \citenamefont {Chiara},\ and\ \citenamefont {Deffner}}]{Myers2022}%
  \BibitemOpen
  \bibfield  {author} {\bibinfo {author} {\bibfnamefont {N.~M.}\ \bibnamefont {Myers}}, \bibinfo {author} {\bibfnamefont {F.~J.}\ \bibnamefont {Peña}}, \bibinfo {author} {\bibfnamefont {O.}~\bibnamefont {Negrete}}, \bibinfo {author} {\bibfnamefont {P.}~\bibnamefont {Vargas}}, \bibinfo {author} {\bibfnamefont {G.~D.}\ \bibnamefont {Chiara}}, \ and\ \bibinfo {author} {\bibfnamefont {S.}~\bibnamefont {Deffner}},\ }\href {\doibase 10.1088/1367-2630/ac47cc} {\bibfield  {journal} {\bibinfo  {journal} {New Journal of Physics}\ }\textbf {\bibinfo {volume} {24}},\ \bibinfo {pages} {025001} (\bibinfo {year} {2022})}\BibitemShut {NoStop}%
\bibitem [{\citenamefont {Estrada}\ \emph {et~al.}(2024)\citenamefont {Estrada}, \citenamefont {Mayo}, \citenamefont {Roncaglia},\ and\ \citenamefont {Mininni}}]{Estrada2024}%
  \BibitemOpen
  \bibfield  {author} {\bibinfo {author} {\bibfnamefont {J.~A.}\ \bibnamefont {Estrada}}, \bibinfo {author} {\bibfnamefont {F.}~\bibnamefont {Mayo}}, \bibinfo {author} {\bibfnamefont {A.~J.}\ \bibnamefont {Roncaglia}}, \ and\ \bibinfo {author} {\bibfnamefont {P.~D.}\ \bibnamefont {Mininni}},\ }\href {\doibase 10.1103/PhysRevA.109.012202} {\bibfield  {journal} {\bibinfo  {journal} {Phys. Rev. A}\ }\textbf {\bibinfo {volume} {109}},\ \bibinfo {pages} {012202} (\bibinfo {year} {2024})}\BibitemShut {NoStop}%
\bibitem [{\citenamefont {Simmons}\ \emph {et~al.}(2023)\citenamefont {Simmons}, \citenamefont {Sajjad}, \citenamefont {Keithley}, \citenamefont {Mas}, \citenamefont {Tanlimco}, \citenamefont {Nolasco-Martinez}, \citenamefont {Bai}, \citenamefont {Fredrickson},\ and\ \citenamefont {Weld}}]{Simmons2023}%
  \BibitemOpen
  \bibfield  {author} {\bibinfo {author} {\bibfnamefont {E.~Q.}\ \bibnamefont {Simmons}}, \bibinfo {author} {\bibfnamefont {R.}~\bibnamefont {Sajjad}}, \bibinfo {author} {\bibfnamefont {K.}~\bibnamefont {Keithley}}, \bibinfo {author} {\bibfnamefont {H.}~\bibnamefont {Mas}}, \bibinfo {author} {\bibfnamefont {J.~L.}\ \bibnamefont {Tanlimco}}, \bibinfo {author} {\bibfnamefont {E.}~\bibnamefont {Nolasco-Martinez}}, \bibinfo {author} {\bibfnamefont {Y.}~\bibnamefont {Bai}}, \bibinfo {author} {\bibfnamefont {G.~H.}\ \bibnamefont {Fredrickson}}, \ and\ \bibinfo {author} {\bibfnamefont {D.~M.}\ \bibnamefont {Weld}},\ }\href {\doibase 10.1103/PhysRevResearch.5.L042009} {\bibfield  {journal} {\bibinfo  {journal} {Phys. Rev. Res.}\ }\textbf {\bibinfo {volume} {5}},\ \bibinfo {pages} {L042009} (\bibinfo {year} {2023})}\BibitemShut {NoStop}%
\bibitem [{\citenamefont {Pethick}\ and\ \citenamefont {Smith}(2008)}]{PethickSmith2008}%
  \BibitemOpen
  \bibfield  {author} {\bibinfo {author} {\bibfnamefont {C.~J.}\ \bibnamefont {Pethick}}\ and\ \bibinfo {author} {\bibfnamefont {H.}~\bibnamefont {Smith}},\ }\href@noop {} {\emph {\bibinfo {title} {Bose–Einstein Condensation in Dilute Gases}}},\ \bibinfo {edition} {2nd}\ ed.\ (\bibinfo  {publisher} {Cambridge University Press},\ \bibinfo {year} {2008})\BibitemShut {NoStop}%
\bibitem [{\citenamefont {Huang}\ \emph {et~al.}(2021)\citenamefont {Huang}, \citenamefont {Modugno},\ and\ \citenamefont {Chen}}]{Huang2021}%
  \BibitemOpen
  \bibfield  {author} {\bibinfo {author} {\bibfnamefont {T.-Y.}\ \bibnamefont {Huang}}, \bibinfo {author} {\bibfnamefont {M.}~\bibnamefont {Modugno}}, \ and\ \bibinfo {author} {\bibfnamefont {X.}~\bibnamefont {Chen}},\ }\href {\doibase 10.1103/PhysRevA.104.063313} {\bibfield  {journal} {\bibinfo  {journal} {Phys. Rev. A}\ }\textbf {\bibinfo {volume} {104}},\ \bibinfo {pages} {063313} (\bibinfo {year} {2021})}\BibitemShut {NoStop}%
\bibitem [{\citenamefont {Huang}\ \emph {et~al.}(2020)\citenamefont {Huang}, \citenamefont {Malomed},\ and\ \citenamefont {Chen}}]{Huang2020}%
  \BibitemOpen
  \bibfield  {author} {\bibinfo {author} {\bibfnamefont {T.-Y.}\ \bibnamefont {Huang}}, \bibinfo {author} {\bibfnamefont {B.~A.}\ \bibnamefont {Malomed}}, \ and\ \bibinfo {author} {\bibfnamefont {X.}~\bibnamefont {Chen}},\ }\href {\doibase 10.1063/5.0004309} {\bibfield  {journal} {\bibinfo  {journal} {Chaos: An Interdisciplinary Journal of Nonlinear Science}\ }\textbf {\bibinfo {volume} {30}},\ \bibinfo {pages} {053131} (\bibinfo {year} {2020})}\BibitemShut {NoStop}%
\bibitem [{\citenamefont {Abah}\ and\ \citenamefont {Lutz}(2018)}]{Abah2018}%
  \BibitemOpen
  \bibfield  {author} {\bibinfo {author} {\bibfnamefont {O.}~\bibnamefont {Abah}}\ and\ \bibinfo {author} {\bibfnamefont {E.}~\bibnamefont {Lutz}},\ }\href {\doibase 10.1103/PhysRevE.98.032121} {\bibfield  {journal} {\bibinfo  {journal} {Phys. Rev. E}\ }\textbf {\bibinfo {volume} {98}},\ \bibinfo {pages} {032121} (\bibinfo {year} {2018})}\BibitemShut {NoStop}%
\bibitem [{\citenamefont {Zhang}\ \emph {et~al.}(2023)\citenamefont {Zhang}, \citenamefont {Yu},\ and\ \citenamefont {Liu}}]{Zhang2023-mx}%
  \BibitemOpen
  \bibfield  {author} {\bibinfo {author} {\bibfnamefont {M.}~\bibnamefont {Zhang}}, \bibinfo {author} {\bibfnamefont {H.-M.}\ \bibnamefont {Yu}}, \ and\ \bibinfo {author} {\bibfnamefont {J.}~\bibnamefont {Liu}},\ }\href@noop {} {\bibfield  {journal} {\bibinfo  {journal} {npj Quantum Information}\ }\textbf {\bibinfo {volume} {9}},\ \bibinfo {pages} {97} (\bibinfo {year} {2023})}\BibitemShut {NoStop}%
\bibitem [{\citenamefont {Abah}\ and\ \citenamefont {Paternostro}(2019)}]{Obinna2019}%
  \BibitemOpen
  \bibfield  {author} {\bibinfo {author} {\bibfnamefont {O.}~\bibnamefont {Abah}}\ and\ \bibinfo {author} {\bibfnamefont {M.}~\bibnamefont {Paternostro}},\ }\href {\doibase 10.1103/PhysRevE.99.022110} {\bibfield  {journal} {\bibinfo  {journal} {Phys. Rev. E}\ }\textbf {\bibinfo {volume} {99}},\ \bibinfo {pages} {022110} (\bibinfo {year} {2019})}\BibitemShut {NoStop}%
\bibitem [{\citenamefont {Abah}\ \emph {et~al.}(2019)\citenamefont {Abah}, \citenamefont {Puebla}, \citenamefont {Kiely}, \citenamefont {Chiara}, \citenamefont {Paternostro},\ and\ \citenamefont {Campbell}}]{Obinna2019_2}%
  \BibitemOpen
  \bibfield  {author} {\bibinfo {author} {\bibfnamefont {O.}~\bibnamefont {Abah}}, \bibinfo {author} {\bibfnamefont {R.}~\bibnamefont {Puebla}}, \bibinfo {author} {\bibfnamefont {A.}~\bibnamefont {Kiely}}, \bibinfo {author} {\bibfnamefont {G.~D.}\ \bibnamefont {Chiara}}, \bibinfo {author} {\bibfnamefont {M.}~\bibnamefont {Paternostro}}, \ and\ \bibinfo {author} {\bibfnamefont {S.}~\bibnamefont {Campbell}},\ }\href {\doibase 10.1088/1367-2630/ab4c8c} {\bibfield  {journal} {\bibinfo  {journal} {New Journal of Physics}\ }\textbf {\bibinfo {volume} {21}},\ \bibinfo {pages} {103048} (\bibinfo {year} {2019})}\BibitemShut {NoStop}%
\bibitem [{\citenamefont {Abah}\ \emph {et~al.}(2020)\citenamefont {Abah}, \citenamefont {Paternostro},\ and\ \citenamefont {Lutz}}]{Obinna2020}%
  \BibitemOpen
  \bibfield  {author} {\bibinfo {author} {\bibfnamefont {O.}~\bibnamefont {Abah}}, \bibinfo {author} {\bibfnamefont {M.}~\bibnamefont {Paternostro}}, \ and\ \bibinfo {author} {\bibfnamefont {E.}~\bibnamefont {Lutz}},\ }\href {\doibase 10.1103/PhysRevResearch.2.023120} {\bibfield  {journal} {\bibinfo  {journal} {Phys. Rev. Res.}\ }\textbf {\bibinfo {volume} {2}},\ \bibinfo {pages} {023120} (\bibinfo {year} {2020})}\BibitemShut {NoStop}%
\bibitem [{\citenamefont {del Campo}\ \emph {et~al.}(2014{\natexlab{b}})\citenamefont {del Campo}, \citenamefont {Goold},\ and\ \citenamefont {Paternostro}}]{Campo2014}%
  \BibitemOpen
  \bibfield  {author} {\bibinfo {author} {\bibfnamefont {A.}~\bibnamefont {del Campo}}, \bibinfo {author} {\bibfnamefont {J.}~\bibnamefont {Goold}}, \ and\ \bibinfo {author} {\bibfnamefont {M.}~\bibnamefont {Paternostro}},\ }\href {\doibase 10.1038/srep06208} {\bibfield  {journal} {\bibinfo  {journal} {Scientific Reports}\ }\textbf {\bibinfo {volume} {4}},\ \bibinfo {pages} {6208} (\bibinfo {year} {2014}{\natexlab{b}})}\BibitemShut {NoStop}%
\bibitem [{\citenamefont {{\c C}akmak}\ and\ \citenamefont {M\"ustecapl\ifmmode \imath \else \i \fi{}o\ifmmode~\breve{g}\else \u{g}\fi{}lu}(2019)}]{Baris_2019}%
  \BibitemOpen
  \bibfield  {author} {\bibinfo {author} {\bibfnamefont {B.}~\bibnamefont {{\c C}akmak}}\ and\ \bibinfo {author} {\bibfnamefont {O.~E.}\ \bibnamefont {M\"ustecapl\ifmmode \imath \else \i \fi{}o\ifmmode~\breve{g}\else \u{g}\fi{}lu}},\ }\href {\doibase 10.1103/PhysRevE.99.032108} {\bibfield  {journal} {\bibinfo  {journal} {Phys. Rev. E}\ }\textbf {\bibinfo {volume} {99}},\ \bibinfo {pages} {032108} (\bibinfo {year} {2019})}\BibitemShut {NoStop}%
\bibitem [{\citenamefont {Hartmann}\ \emph {et~al.}(2020)\citenamefont {Hartmann}, \citenamefont {Mukherjee}, \citenamefont {Niedenzu},\ and\ \citenamefont {Lechner}}]{Hartmann2020}%
  \BibitemOpen
  \bibfield  {author} {\bibinfo {author} {\bibfnamefont {A.}~\bibnamefont {Hartmann}}, \bibinfo {author} {\bibfnamefont {V.}~\bibnamefont {Mukherjee}}, \bibinfo {author} {\bibfnamefont {W.}~\bibnamefont {Niedenzu}}, \ and\ \bibinfo {author} {\bibfnamefont {W.}~\bibnamefont {Lechner}},\ }\href {\doibase 10.1103/PhysRevResearch.2.023145} {\bibfield  {journal} {\bibinfo  {journal} {Phys. Rev. Res.}\ }\textbf {\bibinfo {volume} {2}},\ \bibinfo {pages} {023145} (\bibinfo {year} {2020})}\BibitemShut {NoStop}%
\bibitem [{\citenamefont {Pedram}\ \emph {et~al.}(2023)\citenamefont {Pedram}, \citenamefont {Kadıoğlu}, \citenamefont {Kabak{\c c}ıoğlu},\ and\ \citenamefont {M\"ustecaplıoğlu}}]{Pedram_2023}%
  \BibitemOpen
  \bibfield  {author} {\bibinfo {author} {\bibfnamefont {A.}~\bibnamefont {Pedram}}, \bibinfo {author} {\bibfnamefont {S.~C.}\ \bibnamefont {Kadıoğlu}}, \bibinfo {author} {\bibfnamefont {A.}~\bibnamefont {Kabak{\c c}ıoğlu}}, \ and\ \bibinfo {author} {\bibfnamefont {O.~E.}\ \bibnamefont {M\"ustecaplıoğlu}},\ }\href {\doibase 10.1088/1367-2630/ad0857} {\bibfield  {journal} {\bibinfo  {journal} {New Journal of Physics}\ }\textbf {\bibinfo {volume} {25}},\ \bibinfo {pages} {113014} (\bibinfo {year} {2023})}\BibitemShut {NoStop}%
\bibitem [{\citenamefont {Williamson}\ and\ \citenamefont {Davis}(2024)}]{Lewis2024}%
  \BibitemOpen
  \bibfield  {author} {\bibinfo {author} {\bibfnamefont {L.~A.}\ \bibnamefont {Williamson}}\ and\ \bibinfo {author} {\bibfnamefont {M.~J.}\ \bibnamefont {Davis}},\ }\href {\doibase 10.1103/PhysRevB.109.024310} {\bibfield  {journal} {\bibinfo  {journal} {Phys. Rev. B}\ }\textbf {\bibinfo {volume} {109}},\ \bibinfo {pages} {024310} (\bibinfo {year} {2024})}\BibitemShut {NoStop}%
\bibitem [{\citenamefont {Chen}\ \emph {et~al.}(2019)\citenamefont {Chen}, \citenamefont {Watanabe}, \citenamefont {Yu}, \citenamefont {Guan},\ and\ \citenamefont {del Campo}}]{Chen2019}%
  \BibitemOpen
  \bibfield  {author} {\bibinfo {author} {\bibfnamefont {Y.-Y.}\ \bibnamefont {Chen}}, \bibinfo {author} {\bibfnamefont {G.}~\bibnamefont {Watanabe}}, \bibinfo {author} {\bibfnamefont {Y.-C.}\ \bibnamefont {Yu}}, \bibinfo {author} {\bibfnamefont {X.-W.}\ \bibnamefont {Guan}}, \ and\ \bibinfo {author} {\bibfnamefont {A.}~\bibnamefont {del Campo}},\ }\href {\doibase 10.1038/s41534-019-0204-5} {\bibfield  {journal} {\bibinfo  {journal} {npj Quantum Information}\ }\textbf {\bibinfo {volume} {5}},\ \bibinfo {pages} {88} (\bibinfo {year} {2019})}\BibitemShut {NoStop}%
\bibitem [{\citenamefont {Keller}\ \emph {et~al.}(2020)\citenamefont {Keller}, \citenamefont {Fogarty}, \citenamefont {Li},\ and\ \citenamefont {Busch}}]{Keller2020}%
  \BibitemOpen
  \bibfield  {author} {\bibinfo {author} {\bibfnamefont {T.}~\bibnamefont {Keller}}, \bibinfo {author} {\bibfnamefont {T.}~\bibnamefont {Fogarty}}, \bibinfo {author} {\bibfnamefont {J.}~\bibnamefont {Li}}, \ and\ \bibinfo {author} {\bibfnamefont {T.}~\bibnamefont {Busch}},\ }\href {\doibase 10.1103/PhysRevResearch.2.033335} {\bibfield  {journal} {\bibinfo  {journal} {Phys. Rev. Res.}\ }\textbf {\bibinfo {volume} {2}},\ \bibinfo {pages} {033335} (\bibinfo {year} {2020})}\BibitemShut {NoStop}%
\bibitem [{\citenamefont {Boubakour}\ \emph {et~al.}(2023)\citenamefont {Boubakour}, \citenamefont {Fogarty},\ and\ \citenamefont {Busch}}]{MOMO2023}%
  \BibitemOpen
  \bibfield  {author} {\bibinfo {author} {\bibfnamefont {M.}~\bibnamefont {Boubakour}}, \bibinfo {author} {\bibfnamefont {T.}~\bibnamefont {Fogarty}}, \ and\ \bibinfo {author} {\bibfnamefont {T.}~\bibnamefont {Busch}},\ }\href {\doibase 10.1103/PhysRevResearch.5.013088} {\bibfield  {journal} {\bibinfo  {journal} {Phys. Rev. Res.}\ }\textbf {\bibinfo {volume} {5}},\ \bibinfo {pages} {013088} (\bibinfo {year} {2023})}\BibitemShut {NoStop}%
\bibitem [{\citenamefont {Watson}\ and\ \citenamefont {Kheruntsyan}(2023)}]{Watson2023}%
  \BibitemOpen
  \bibfield  {author} {\bibinfo {author} {\bibfnamefont {R.~S.}\ \bibnamefont {Watson}}\ and\ \bibinfo {author} {\bibfnamefont {K.~V.}\ \bibnamefont {Kheruntsyan}},\ }\href@noop {} {\enquote {\bibinfo {title} {Quantum many-body thermal machines enabled by atom-atom correlations},}\ } (\bibinfo {year} {2023}),\ \Eprint {http://arxiv.org/abs/arXiv:2308.05266} {arXiv:2308.05266} \BibitemShut {NoStop}%
\bibitem [{\citenamefont {Koch}\ \emph {et~al.}(2023)\citenamefont {Koch}, \citenamefont {Menon}, \citenamefont {Cuestas}, \citenamefont {Barbosa}, \citenamefont {Lutz}, \citenamefont {Fogarty}, \citenamefont {Busch},\ and\ \citenamefont {Widera}}]{Koch2023}%
  \BibitemOpen
  \bibfield  {author} {\bibinfo {author} {\bibfnamefont {J.}~\bibnamefont {Koch}}, \bibinfo {author} {\bibfnamefont {K.}~\bibnamefont {Menon}}, \bibinfo {author} {\bibfnamefont {E.}~\bibnamefont {Cuestas}}, \bibinfo {author} {\bibfnamefont {S.}~\bibnamefont {Barbosa}}, \bibinfo {author} {\bibfnamefont {E.}~\bibnamefont {Lutz}}, \bibinfo {author} {\bibfnamefont {T.}~\bibnamefont {Fogarty}}, \bibinfo {author} {\bibfnamefont {T.}~\bibnamefont {Busch}}, \ and\ \bibinfo {author} {\bibfnamefont {A.}~\bibnamefont {Widera}},\ }\href {\doibase 10.1038/s41586-023-06469-8} {\bibfield  {journal} {\bibinfo  {journal} {Nature}\ }\textbf {\bibinfo {volume} {621}},\ \bibinfo {pages} {723} (\bibinfo {year} {2023})}\BibitemShut {NoStop}%
\end{thebibliography}%
\end{document}